\documentclass[reprint,showpacs,groupedaddress,eqsecnum,amsfonts,amsmath,amssymb,aps,prd]{revtex4}
\usepackage{graphicx}
\usepackage{latexsym}
\usepackage{amssymb}
\usepackage{mathrsfs}
\usepackage{amsmath}
\usepackage{amsthm}
\usepackage{color}
\usepackage{dcolumn}
\usepackage{bm}

\definecolor{dyellow}{rgb}{1.,0.8,.0}
\definecolor{myblue}{rgb}{.1,.1,.7}
\definecolor{dcyan}{rgb}{.0,.6,.6}
\definecolor{dmagenta}{rgb}{0.6,0.0,0.6}
\definecolor{brown}{rgb}{0.6,0.2,0.}
\definecolor{darkblue}{rgb}{.0,.0,0.5}
\definecolor{darkred}{rgb}{0.75,0.0,0.0}
\definecolor{orange}{rgb}{1.,.6,.0}
\definecolor{dorange}{rgb}{0.8,.4,.0}
\definecolor{darkgreen}{rgb}{0.0,0.6,0.0}
\definecolor{purple}{rgb}{.4,.0,.4}
\definecolor{grey}{rgb}{0.5,0.5,0.5}

\begin{document}
\hyphenpenalty=1000
\preprint{APS/123-QED}
\title{Multipole analysis in the radiation field for linearized $f(R)$ gravity with irreducible Cartesian tensors}

\newcommand*{\PKU}{Institute of High Energy Physics and Theoretical Physics Center for Science Facilities,
Chinese Academy of Sciences, Beijing, 100049, People's Republic of China}\affiliation{\PKU}
\newcommand*{\INFN}{INFN, Sez. di Pavia, via Bassi 6, 27100 Pavia, Italy}\affiliation{\INFN}
\newcommand*{\CICQM}{}\affiliation{\CICQM}
\newcommand*{\CHEP}{}\affiliation{\CHEP}

\author{Bofeng Wu}\email{wubf@ihep.ac.cn}\affiliation{\PKU}
\author{Chao-Guang Huang}\email{huangcg@ihep.ac.cn}\affiliation{\PKU}

\begin{abstract}
The $1/r$-expansion in the distance to the source is applied to the linearized $f(R)$ gravity, and its multipole expansion in the radiation field with irreducible Cartesian tensors is presented.
Then, the energy, momentum, and angular momentum in the gravitational waves are provided for linearized $f(R)$ gravity. All of these results have two parts which are associated with the tensor part and the scalar part in the multipole expansion of linearized $f(R)$ gravity, respectively. The former is the same as that in General Relativity, and the latter, as the correction to the result in General Relativity, is caused by the massive scalar degree of freedom, and places an important role in distinguishing GR and $f(R)$ gravity.
\end{abstract}
\pacs{04.50.Kd,  04.30.-w, 04.25.Nx}
\maketitle
\section{Introduction}
General Relativity (GR) is a great and successful theory of gravity. The detection of gravitational waves (GWs) by the LIGO and Virgo Collaboration~\cite{TheLIGOScientific:2016agk}
is a milestone in GWs and shows that the observations of GWs are consistent very well with GR's prediction, which
further promotes the study of GR and astrophysics~\cite{TheLIGOScientific:2016src,TheLIGOScientific:2016htt,GBM:2017lvd}.

In despite of the above great success, there are still many challenges for GR to face, \emph{e.g.}, interpreting many data observed at infrared scales~\cite{AstierP2006,Eisenstein2005,Riess2004,Spergel2007}. Introducing the modified gravity theory~\cite{Sotiriou2010} is one way to deal with these difficulties.
$f(R)$ gravity~\cite{Buchdahl1970,Starobinsky:1980te,York1972,Gibbons1977,Nojiri:2010wj,DNojiri:2017ncd} is the most typical example of the modified gravity theory, and it replaces the Einstein-Hilbert action by the quantity $f(R)$ in the gravitational Lagrangian, where $f$ is a general function of the Ricci scalar $R$.

For the sources localized in a finite region of space, the multipole expansion is one of the most convenient and useful ways of describing the external field~\cite{Damour:1990gj}. One of the important method with respect to the multipole expansion is the symmetric and trace-free (STF) formalism in terms of the irreducible Cartesian tensors, which is developed by Thorne, Blanchet, Damour and
Iyer~\cite{Thorne:1980ru,Blanchet:1985sp,Blanchet:1989ki}. The STF technique is summarized in Ref.~\cite{Damour:1990gj}. The GW possesses more polarizations for the modified gravity theory~\cite{Rizwana:2016qdq}, so it is worthwhile exploring the radiation field in the modified gravity theory, especially in $f(R)$ gravity, in terms of the STF formalism, for it will contribute to our understanding the possible different polarizations of GWs~\cite{Wu:2017huang}.

In our preceding paper~\cite{Wu:2017huang}, referred to as I~\label{I} hereafter, the field equations of $f(R)$ gravity are rewritten in the form of obvious wave equations with the stress-energy pseudotensor of the matter fields and the gravitational field as  the sources under de Donder condition, which are similar to those of GR~\cite{Thorne:1980ru,Blanchet:2013haa}. Then, the linearized $f(R)$ gravity is presented, the STF method is applied to the linearized $f(R)$ gravity, and the explicit expressions of multipole expansion are derived. In these expressions, the tensor part is associated with the effective gravitational field amplitude of $f(R)$ gravity
\begin{equation}\label{equ1.1}
\tilde{h}^{\mu\nu}:=f_{R}\sqrt{-g}g^{\mu\nu}-\eta^{\mu\nu},
\end{equation}
and the scalar part is associated with the linear part of Ricci scalar $R^{(1)}$, where $g^{\mu\nu}$ denotes the contravariant metric, $\eta^{\mu\nu}$ represents an auxiliary Minkowskian metric, $g$ is the determinant of metric $g_{\mu\nu}$, and $f_{R}=\partial_{R}f$.

In this paper, we shall deal with the multipole expansion of linearized $f(R)$ gravity in the radiation field with irreducible Cartesian tensors, which requires that the $1/r$-expansion in the distance to the source is applied to the linearized $f(R)$ gravity. This is feasible by the STF technique, because the multipole expansion of linearized $f(R)$ gravity has been expressed in terms of the STF formalism \cite{Wu:2017huang}. In the multipole expansion of the tensor part, the multipole moments of the source are only the functions of
the retarded time $u=t-r/c$, as in GR, where $r$ is the distance to the source, and $c$ is the velocity of light in vacuum. Whereas in the multipole expansion of the scalar part~\cite{Wu:2017huang,Naf:2011za}, the multipole moments of the source are the functions of both $t$ and $r$ instead of the retarded time $u$.

Based on the above results, the energy, momentum, and angular momentum in the GWs for linearized $f(R)$ gravity are presented. For the linearized GR, the energy and the momentum carried by GWs can be derived directly by use of its effective stress-energy of GWs, which is the average of quadratic terms of gravitational perturbation over a small volume for several wavelengths~\cite{Berry:2011pb,Carroll2004}.
However, the angular momentum cannot be derived in this way~\cite{Thorne:1980ru}.  The similar difficulty also appears in the linearized $f(R)$ gravity. In this paper, by following Refs.~\cite{Thorne:1980ru,Peters:1964zz}, we shall deal with the energy, momentum, and angular momentum in the GWs for linearized $f(R)$ gravity in a unified way.

The expressions of energy, momentum, and angular momentum in the GWs for linearized $f(R)$ gravity have also two parts which are associated with the tensor part and the scalar part in the multipole expansion of linearized $f(R)$ gravity, respectively. We shall show that the former is the same as that in GR, and the latter, as the correction to the result in GR, results from the massive scalar degree of freedom in linearized $f(R)$ gravity. It is very these corrections that show the monopole and the dipole radiations appear in $f(R)$ gravity, as in literature~\cite{Naf:2011za}. Further, they show that the monopole radiation plays the key role in distinguishing GR and $f(R)$ gravity.

This paper is organized as follows. In Sec.~\ref{Sec:Preliminary}, we show the notation, the
relevant formulas of STF formalism, and the review of the metric $f(R)$ gravity.
In Sec.~\ref{Sec:MultipoleExpansion}, based on the result about the multipole expansion of linearized $f(R)$ gravity in~\ref{I}, we derive its corresponding multipole expansion in the radiation field by using the $1/r$-expansion in the distance to the source. In Sec.~\ref{Sec:EneggyMomentumAgular}, we evaluate the energy, momentum, and angular momentum in the GWs for linearized $f(R)$ gravity systematically. In
Sec.~\ref{Sec:Conclusion}, we present the conclusions and make some discussions.
In Appendix, we provide the detailed derivation about (\ref{equ4.29}).
\section{Preliminary\label{Sec:Preliminary}}
\subsection{Notation~\label{Sec:Notation}}
The notation in this paper is the same as in~\ref{I}. The international system of units is used throughout, and the metric $g^{\mu\nu}$ has signature $(-,+,+,+)$. The Greek indices represent
spacetime indices and run from 0 to 3. They follow the Einstein summation rule. The Latin indices represent spatial indices and run from 1 to 3. They follow the rule that the
repeated Latin indices are to be summed as though a
$\delta_{ij}$ was present. The 3-dimensional Levi-Civita symbol is denoted by $\epsilon_{ijk}$ with $\epsilon_{123}=1$.
When the linearized gravitational theory is discussed,
$(x^{\mu})=(ct,x_{i})$,
as though they were Minkowskian coordinates.
The Cartesian coordinates define the spherical coordinate system $(ct,r,\theta,\varphi)$:
\begin{equation}\label{equ2.1}
x_{1}=r\sin{\theta}\cos{\varphi},\ x_{2}=r\sin{\theta}\sin{\varphi},\ x_{3}=r\cos{\theta}.
\end{equation}

As in a flat space, the radial vector and its length are denoted by $\boldsymbol{x}$ and $r$, respectively.
The unit radial vector is $\boldsymbol{n}$, and its components are $n_{i}$, so that $n_{i}=x_{i}/r$, where $x_{i}$
are the components of $\boldsymbol{x}$. Obviously, from (\ref{equ2.1}), there is
\begin{equation}\label{equ2.2}
\partial_{r}=\frac{\partial}{\partial r}=n_{1}\frac{\partial}{\partial x_{1}}+n_{2}\frac{\partial}{\partial x_{2}}+n_{3}\frac{\partial}{\partial x_{3}}=n_{i}\partial_{i}.
\end{equation}

A Cartesian tensor with $l$ indices is denoted by~\cite{Thorne:1980ru}
\begin{equation}\label{equ2.3}
B_{I_{l}}\equiv B_{i_{1}i_{2}\cdots i_{l}},
\end{equation}
and especially, the tensor products of $l$ radial and unit radial vectors are abbreviated by
\begin{align}
\label{equ2.4}X_{I_{l}}=X_{i_{1}i_{2}\cdots i_{l}}:= x_{i_{1}}x_{i_{2}}\cdots x_{i_{l}},\\
\label{equ2.5}N_{I_{l}}=N_{i_{1}i_{2}\cdots i_{l}}:= n_{i_{1}}n_{i_{2}}\cdots n_{i_{l}}
\end{align}
with
\begin{equation}\label{equ2.6}
X_{I_{l}}=r^l N_{I_{l}}.
\end{equation}

\subsection{The relevant formulas in STF formalism~\cite{Damour:1990gj,Thorne:1980ru,Blanchet:1985sp,Wu:2017huang}~\label{Sec:STFformulae}}
In this subsection, the relevant formulas in STF formalism are listed.
The symmetric part of a Cartesian tensor $B_{I_{l}}$ is expressed by
\begin{equation}\label{equ2.7}
B_{(I_{l})}=B_{(i_{1}i_{2}\cdots i_{l})}:=\frac{1}{l!}\sum_{\sigma} B_{i_{\sigma(1)}i_{\sigma(2)}\cdots i_{\sigma(l)}},
\end{equation}
where $\sigma$ runs over all permutations of $(12\cdots l)$. The STF part of $B_{I_{l}}$ is
\begin{align}
\label{equ2.8}
\hat{B}_{I_{l}}&\equiv B_{<I_{l}>}\equiv B_{<i_{1}i_{2}\cdots i_{l}>}
:=\sum_{k=0}^{[\frac{l}{2}]}b_{k}\delta_{(i_{1}i_{2}}\cdots\delta_{i_{2k-1}i_{2k}}S_{i_{2k+1}\cdots i_{l})a_{1}a_{1}\cdots a_{k}a_{k}},\quad
\end{align}
where
\begin{align}
\label{equ2.9}S_{I_{l}}=&B_{(I_{l})},\\
\label{equ2.10}b_{k}=&(-1)^{k}\frac{(2l-2k-1)!!}{(2l-1)!!}\frac{l!}{(2k)!!(l-2k)!}.
\end{align}

By the above formulas, there are
\begin{align}
\label{equ2.11}\hat{N}_{I_{l}}&=\sum_{k=0}^{[\frac{l}{2}]}b_{k}\delta_{(i_{1}i_{2}}\cdots\delta_{i_{2k-1}i_{2k}}
N_{i_{2k+1}\cdots i_{l})},\\
\label{equ2.12}\hat{\partial}_{I_{l}}&=\sum_{k=0}^{[\frac{l}{2}]}b_{k}\delta_{(i_{1}i_{2}}\cdots\delta_{i_{2k-1}i_{2k}}
\partial_{i_{2k+1}\cdots i_{l})},\\
\label{equ2.13}\partial_{i}n_{j}&=\frac{1}{r}(\delta_{ij}-n_{i}n_{j}),\\
\label{equ2.14}\partial_{i}\left(F\left(t-\frac{\epsilon r}{c}\right)\right)&=-\frac{\epsilon n_{i}}{c}\partial_{t}F\left(t-\frac{\epsilon r}{c}\right),\qquad (\epsilon^{2}=1),\\
\label{equ2.15}\hat{\partial}_{I_{l}}\left(\frac{F(t-\epsilon r/c)}{r}\right)&=(-\epsilon)^{l}\hat{N}_{I_{l}}\sum_{k=0}^{l}\frac{(l+k)!}{(2\epsilon)^{k}k!(l-k)!}
\frac{\partial_{t}^{l-k}F(t-\epsilon r/c)}{c^{l-k}r^{k+1}},\qquad (\epsilon^{2}=1),
\end{align}
where $\partial_{I_{l}}\equiv\partial_{i_{1}i_{2}\cdots i_{l}}:=\partial_{i_{1}}\partial_{i_{2}}\cdots\partial_{i_{l}}$,
and $\partial_t^{l-k}$ is the $(l-k)^{\rm th}$ derivative with respect to $t$.

The most useful formulas about integrals over a unit sphere are
\begin{align}
\label{equ2.16}&\int N_{I_{2l+1}}d\Omega=0,\\
\label{equ2.17}&\int N_{I_{2l}}d\Omega=\frac{4\pi}{2l+1}\delta_{(i_{1}i_{2}}\cdots\delta_{i_{2l-1}i_{2l})},
\end{align}
where $l$ is the nonnegtive integer, and $\Omega$ is the solid angle about the radial vector. Then, for any two STF tensors $\hat{A}_{I_{l}}$ and $\hat{B}_{J_{l'}}$, we have
\begin{align}
\label{equ2.18}\int\Big(\hat{A}_{I_{l}}N_{I_{l}}\Big)\Big(\hat{B}_{J_{l'}}N_{J_{l'}}\Big)d\Omega&=\frac{ 4\pi l!}{(2l+1)!!}\hat{A}_{I_{l}}\hat{B}_{I_{l}}\delta_{l\,l'},\\
\label{equ2.19}\int\Big(\hat{A}_{I_{l}}N_{I_{l}}\Big)\Big(\hat{B}_{J_{l'}}N_{J_{l'}}\Big)n_{k}d\Omega&=4\pi\left(\frac{ (l'+1)!}{(2l'+3)!!}\hat{A}_{kJ_{l'}}\hat{B}_{J_{l'}}\delta_{l\,l'+1}+\frac{ (l+1)!}{(2l+3)!!}\hat{A}_{I_{l}}\hat{B}_{kI_{l}}\delta_{l'\,l+1}\right).
\end{align}

\subsection{Review of the metric $f(R)$ gravity~\cite{Wu:2017huang}~\label{Sec:f(R)gravity}}
We restrict our attention to the metric $f(R)$ gravity, and its action is
\begin{equation}\label{equ2.20}
S=\frac{1}{2\kappa c}\int dx^4\sqrt{-g}f(R)+S_{M}(g^{\mu\nu},\psi),
\end{equation}
where $f$ is an arbitrary function of Ricci scalar $R$, $\kappa=8\pi G/c^{4}$, and $S_{M}(g^{\mu\nu},\psi)$ is the matter action. Varying this action with respect to the metric $g^{\mu\nu}$ gives the gravitational field equations
\begin{equation}\label{equ2.21}
H_{\mu\nu}=\kappa T_{\mu\nu},
\end{equation}
where
\begin{align}
\label{equ2.22}H_{\mu\nu}&=-\frac{g_{\mu\nu}}{2}f+(R_{\mu\nu}+g_{\mu\nu}\square-\nabla_{\mu}\nabla_{\nu})f_{R}
\end{align}
and $T_{\mu\nu}$
is the stress-energy tensor of matter fields. We only consider polynomial $f(R)$ models of the form
\begin{equation}\label{equ2.23}
f(R)=R+a R^{2}+b R^{3}+\cdots,
\end{equation}
where $a,b\cdots$ are the coupling constants, and their dimensions are $[R]^{-1},[R]^{-2}\cdots$, respectively.


\section{The multipole expansion of linearized $f(R)$ gravity in the radiation field~\label{Sec:MultipoleExpansion}}
In~\ref{I},  the field equations of $f(R)$ gravity are rewritten
in the form of obvious wave equations with the stress-energy pseudotensor of the matter fields and the gravitational field as its source under de Donder condition, the linearized $f(R)$ gravity is obtained by the post-Minkowskian method, and the explicit expressions of multipole expansion {for linearized $f(R)$ gravity are derived. The following is a brief review.

In $f(R)$ gravity, the tensor part is described by the effective gravitational amplitude $\tilde{h}^{\mu\nu}$ which is defined by \eqref{equ1.1}~\cite{Wu:2017huang}.
It contains, besides the information of the metric, }the  information of the function $f_{R}$
which is, from (\ref{equ2.23}),
\begin{equation}\label{equ3.1}
f_{R}=1+2a R+3b R^{2}+\cdots.
\end{equation}
The effective gravitational amplitude $\tilde{h}^{\mu\nu}$ is related to the gravitational amplitude
\begin{align}
\label{equ3.2}h^{\mu\nu}&:=\sqrt{-g}g^{\mu\nu}-\eta^{\mu\nu}
\end{align}
by
\begin{align}
\label{equ3.3}h^{\mu\nu}&=\frac{1}{f_{R}}\tilde{h}^{\mu\nu}+\left(\frac{1}{f_{R}}-1\right)\eta^{\mu\nu}.
\end{align}
Under the de Donder condition,
\begin{equation}\label{equ3.4}
\partial_{\mu}\tilde{h}^{\mu\nu}=0,
\end{equation}
the field equations (\ref{equ2.21}) of $f(R)$ gravity can be transformed into the form of obvious wave equations
\begin{equation}\label{equ3.5}
\square_{\eta} \tilde{h}^{\mu\nu}=2\kappa\tau^{\mu\nu}_{f},
\end{equation}
where $\square_{\eta}:=\eta^{\mu\nu}\partial_{\mu}\partial_{\nu}$, and the source term
\begin{equation}\label{equ3.6}
\tau^{\mu\nu}_{f}=|g|f_{R}^{2}T^{\mu\nu}+\frac{1}{2\kappa}\Lambda^{\mu\nu}_{f}
\end{equation}
is the stress-energy pseudotensor of the matter fields and the gravitational field. Here,
\begin{align}
\label{equ3.7}\Lambda^{\mu\nu}_{f}=&-\tilde{h}^{\alpha\beta}\partial_{\alpha}\partial_{\beta}\tilde{h}^{\mu\nu}
-(f_{R}-1)\tilde{g}^{\alpha\beta}\partial_{\alpha}\tilde{h}^{\mu\nu}\partial_{\beta}\ln{f_{R}}
-\frac{1}{2}(1+2f_{R})\tilde{g}^{\mu\nu}\tilde{g}^{\alpha\beta}\partial_{\alpha}\ln{f_{R}}\partial_{\beta}\ln{f_{R}}\notag\\
&-(1-4f_{R})\tilde{g}^{\mu\alpha}\tilde{g}^{\nu\beta}\partial_{\alpha}\ln{f_{R}}\partial_{\beta}\ln{f_{R}}
-2(f_{R}-1)\tilde{g}^{\mu\nu}\tilde{g}^{\alpha\beta}\partial_{\alpha}\partial_{\beta}\ln{f_{R}}
+2(f_{R}-1)\tilde{g}^{\mu\alpha}\tilde{g}^{\nu\beta}\partial_{\alpha}\partial_{\beta}\ln{f_{R}}\notag\\
&-2\tilde{g}_{\beta\tau}\tilde{g}^{\alpha(\mu}\partial_{\lambda}\tilde{h}^{\nu)\tau}\partial_{\alpha}\tilde{h}^{\beta\lambda}
-(f_{R}-1)\tilde{g}_{\rho\sigma}\tilde{g}^{\mu(\alpha}\tilde{g}^{\beta)\nu}\partial_{\alpha}\tilde{h}^{\rho\sigma}\partial_{\beta}\ln{f_{R}}
+\tilde{g}_{\alpha\beta}\tilde{g}^{\lambda\tau}\partial_{\lambda}\tilde{h}^{\mu\alpha}\partial_{\tau}\tilde{h}^{\nu\beta}\notag\\
&+\partial_{\alpha}\tilde{h}^{\mu\beta}\partial_{\beta}\tilde{h}^{\nu\alpha}
-2(1-f_{R})\tilde{g}^{\alpha(\mu}\partial_{\alpha}\tilde{h}^{\nu)\beta}\partial_{\beta}\ln{f_{R}}
-\frac{1}{2}(1-f_{R})\tilde{g}_{\rho\sigma}\tilde{g}^{\mu\nu}\tilde{g}^{\alpha\beta}\partial_{\alpha}\tilde{h}^{\rho\sigma}\partial_{\beta}\ln{f_{R}}\notag\\
&+\frac{1}{2}\tilde{g}_{\alpha\beta}\tilde{g}^{\mu\nu}\partial_{\lambda}\tilde{h}^{\alpha\tau}\partial_{\tau}\tilde{h}^{\beta\lambda}
+\frac{1}{8}(2\tilde{g}^{\mu\alpha}\tilde{g}^{\nu\beta}
-\tilde{g}^{\mu\nu}\tilde{g}^{\alpha\beta})(2\tilde{g}_{\lambda\tau}\tilde{g}_{\epsilon\pi}-\tilde{g}_{\epsilon\tau}\tilde{g}_{\lambda\pi})
\partial_{\alpha}\tilde{h}^{\lambda\pi}\partial_{\beta}\tilde{h}^{\tau\epsilon}\notag\\
&+a\sqrt{-g}f_{R}\tilde{g}^{\mu\nu}R^{2}+b\sqrt{-g}f_{R}\tilde{g}^{\mu\nu}R^{3}+4agf_{R}^{2}R^{\mu\nu}R+6bgf_{R}^{2}R^{\mu\nu}R^{2}+\mbox{higher order terms},
\end{align}
where
\begin{align}
\label{equ3.8}\tilde{g}_{\mu\nu}:=\frac{1}{\sqrt{-g}f_{R}}g_{\mu\nu},\quad
\tilde{g}^{\mu\nu}:=\sqrt{-g}f_{R}g^{\mu\nu}.
\end{align}
As expected, $\Lambda^{\mu\nu}_{f}$ is made of, at least, quadratic terms in the effective gravitational field amplitude $\tilde{h}^{\mu\nu}$ and its first and second derivatives. Obviously, Eqs.~(\ref{equ3.4}) and (\ref{equ3.5}) imply the conservation of the stress-energy pseudotensor in $f(R)$ gravity:
\begin{equation}
\label{equ3.9}\partial_{\mu}\tau^{\mu\nu}_{f}=0.
\end{equation}

If $\tilde{h}^{\mu\nu}$ is a perturbation, namely,
\begin{equation}\label{equ3.10}
|\tilde{h}^{\mu\nu}|\ll1,
\end{equation}
the linearized gravitational field equations reduce to
\begin{align}
\label{equ3.11}\square_{\eta}\tilde{h}&^{\mu\nu}=2\kappa T^{\mu\nu},
\end{align}
and the quadratic term of the stress-energy pseudotensor in the radiation field is
\begin{align}\label{equ3.12}
(\tau^{\mu\nu}_{f})^{(2)}_{\rm{rad}}&=\frac{1}{2\kappa}\bigg(-\tilde{h}^{\alpha\beta}\partial_{\alpha}\partial_{\beta}\tilde{h}^{\mu\nu}
+\frac{1}{2}\partial^{\mu}\tilde{h}_{\alpha\beta}\partial^{\nu}\tilde{h}^{\alpha\beta}
-\frac{1}{4}\partial^{\mu}\tilde{h}\partial^{\nu}\tilde{h}
+\partial_{\beta}\tilde{h}^{\mu\alpha}(\partial^{\beta}\tilde{h}^{\nu}_{\alpha}+\partial_{\alpha}\tilde{h}^{\nu\beta})
-\partial^{\mu}\tilde{h}_{\alpha\beta}\partial^{\alpha}\tilde{h}^{\nu\beta}\notag\\
&-\partial^{\nu}\tilde{h}_{\alpha\beta}\partial^{\alpha}\tilde{h}^{\mu\beta}+\eta^{\mu\nu}\Big(-\frac{1}{4}\partial_{\lambda}\tilde{h}_{\alpha\beta}\partial^{\lambda}\tilde{h}^{\alpha\beta}
+\frac{1}{8}\partial_{\lambda}\tilde{h}\partial^{\lambda}\tilde{h}
+\frac{1}{2}\partial_{\alpha}\tilde{h}_{\beta\lambda}\partial^{\beta}\tilde{h}^{\alpha\lambda}\Big)
+12a^{2}\partial^{\mu}R^{(1)}\partial^{\nu}R^{(1)}\notag \\
&-\eta^{\mu\nu}\Big(a\Big(R^{(1)}\Big)^2+6a^2\partial^{\alpha}R^{(1)}\partial_{\alpha}R^{(1)}\Big)\bigg),
\end{align}
where $\partial^{\mu}=\eta^{\mu\lambda}\partial_\lambda$, $\tilde{h}=\eta_{\mu\nu}\tilde{h}^{\mu\nu}$, $R^{(1)}$ satisfies a massive Klein-Gordon equation with an external source
\begin{align}
\label{equ3.13}\square_{\eta} R^{(1)}-m^{2}R^{(1)}=m^{2}\kappa T^{(1)}
\end{align}
with
\begin{align}
\label{equ3.14}T=g^{\mu\nu}T_{\mu\nu},\quad m^{2}:=\frac{1}{6a}.
\end{align}
The superscripts $(1)$ and $(2)$ represent the linear and quadratic part for a quantity, respectively.
Under the condition (\ref{equ3.10}), the relationship (\ref{equ3.3}) between $h^{\mu\nu}$ and $\tilde{h}^{\mu\nu}$ reduces to
\begin{equation}
\label{equ3.15}h^{\mu\nu}=\tilde{h}^{\mu\nu}-2aR^{(1)}\eta^{\mu\nu}
\end{equation}
which shows that $h^{\mu\nu}$ contains the tensor part $\tilde{h}^{\mu\nu}$ and the scalar part associated with $R^{(1)}$.

Eqs.~(\ref{equ3.11}) and (\ref{equ3.4}) show that $\tilde{h}^{\mu\nu}$ in linearized $f(R)$ gravity and $h^{\mu\nu}$ in linearized GR satisfy the same wave equations and the same gauge condition. So, the multipole expansion of $\tilde{h}^{\mu\nu}$ in linearized $f(R)$ gravity are the same as that of $h^{\mu\nu}$ in linearized GR, namely~\cite{Blanchet:1985sp,Damour:1990gj},
\begin{equation}\label{equ3.16}
\left\{\begin{array}{l}
\displaystyle\tilde{h}^{00}(t,\boldsymbol{x})= -\frac{4G}{c^{2}}\sum_{l=0}^{\infty}\frac{(-1)^{l}}{l!}\partial_{I_{l}}\left( \frac{\hat{M}_{I_{l}}(u)}{r}\right),\smallskip\\
\displaystyle\tilde{h}^{0i}(t,\boldsymbol{x})= \frac{4G}{c^{3}}\sum_{l=1}^{\infty}\frac{(-1)^{l}}{l!}\partial_{I_{l-1}}\left(\frac{\partial_{t}\hat{M}_{iI_{l-1}}(u)}{r}\right) +\frac{4G}{c^{3}}\sum_{l=1}^{\infty}\frac{(-1)^{l}l}{(l+1)!}\epsilon_{iab}\partial_{aI_{l-1}}\left( \frac{\hat{S}_{bI_{l-1}}(u)}{r}\right),\smallskip\\
\displaystyle\tilde{h}^{ij}(t,\boldsymbol{x})=-\frac{4G}{c^{4}}\sum_{l=2}^{\infty}\frac{(-1)^{l}}{l!}\partial_{I_{l-2}}\left(\frac{\partial_{t}^{2}\hat{M}_{ijI_{l-2}}(u)}{r}\right) -\frac{8G}{c^{4}}\sum_{l=2}^{\infty}\frac{(-1)^{l}l}{(l+1)!}\partial_{aI_{l-2}}\left(\frac{\epsilon_{ab(i}\partial_{|t|}\hat{S}_{j)bI_{l-2}}(u)}{r}\right),
\end{array}\right.
\end{equation}
where
\begin{equation}\label{equ3.17}
\left\{\begin{array}{l}
\displaystyle\hat{M}_{I_{l}}(u)=\frac{1}{c^{2}}\int d^{3}x'\left( \hat{X'}_{I_{l}}\left(\overline{T}^{00}_{l}(u,\boldsymbol{x}')+\overline{T}^{aa}_{l}(u,\boldsymbol{x}')\right)
-\frac{4(2l+1)}{c(l+1)(2l+3)}\hat{X'}_{aI_{l}}\partial_{t}\overline{T}^{0a}_{l+1}(u,\boldsymbol{x}') \right . \\
\displaystyle\qquad\qquad\qquad\qquad\qquad\qquad\qquad\left .
+\frac{2(2l+1)}{c^2(l+1)(l+2)(2l+5)}\hat{X'}_{abI_{l}}\partial_{t}^{2}\overline{T}^{ab}_{l+2}(u,\boldsymbol{x}')\right ),\smallskip\\
\displaystyle\hat{S}_{I_{l}}(u)=\frac{1}{c}\int d^{3}x'\left(\epsilon_{ab<i_{1}}\hat{X'}_{|a|i_{2}\cdots i_{l}>}\overline{T}^{0b}_{l}(u,\boldsymbol{x}') \right . \\
\displaystyle\qquad\qquad\qquad\qquad\qquad\qquad\qquad\left . -\frac{2l+1}{c(l+2)(2l+3)}\epsilon_{ab<i_{1}}\hat{X'}_{|ac|i_{2}\cdots i_{l}>}\partial_{t}\overline{T}^{cb}_{l+1}(u,\boldsymbol{x}')\right),\ l\geq1
\end{array}\right.
\end{equation}
are referred to as the mass-type and current-type source multipole moments, respectively~\cite{Blanchet:2013haa},
the symbol $(i|t|j)$ represents that $t$ is not the symmetric indices, the symbol $<i_{1}|a|i_{2}\cdots i_{l}>$ and $<i_{1}|ac|i_{2}\cdots i_{l}>$
represent that $a$ and $ac$ are not STF indices, respectively, and~\cite{Damour:1990gj}
\begin{align}
\label{equ3.18}\overline{T}^{\mu\nu}_{l}(u,\boldsymbol{x}'):=\frac{(2l+1)!!}{2^{l+1}l!}\int_{-1}^{1}(1-z^2)^{l}T^{\mu\nu}\Big(u+\frac{zr'}{c},
\boldsymbol{x}'\Big)dz.
\end{align}

The further gauge can be chosen to simplify the equations in (\ref{equ3.16}). Under an infinitesimal coordinate transformation, $x'^{\mu}\rightarrow x^{\mu}+\varepsilon^{\mu}$,
\begin{equation}
\label{equ3.19}h'^{\mu\nu}=h^{\mu\nu}+\partial^{\mu}\varepsilon^{\nu}+\partial^{\nu}\varepsilon^{\mu}-\eta^{\mu\nu}\partial_{\alpha}\varepsilon^{\alpha}. \end{equation}
Moreover, $R^{(1)}$ is an invariant under the coordinate transformation, so from (\ref{equ3.15}),
\begin{equation}
\label{equ3.20}\tilde{h}'^{\mu\nu}=\tilde{h}^{\mu\nu}+\partial^{\mu}\varepsilon^{\nu}+\partial^{\nu}\varepsilon^{\mu}-\eta^{\mu\nu}
\partial_{\alpha}\varepsilon^{\alpha},
\end{equation}
which shows that the transverse and traceless (TT) gauge can be applied to the effective gravitational field amplitude $\tilde{h}^{\mu\nu}$ like that in GR~\cite{Naf:2011za}. Hence, under the TT gauge, (\ref{equ3.16}) can be written as
\begin{equation}\label{equ3.21}
\left\{\begin{array}{l}
\displaystyle\tilde{h}^{00}_{TT}(t,\boldsymbol{x})= 0,\bigskip\\
\displaystyle\tilde{h}^{0i}_{TT}(t,\boldsymbol{x})= 0,\medskip\\
\displaystyle\tilde{h}^{ij}_{TT}(t,\boldsymbol{x})=-\frac{4G}{c^{4}}\left(\sum_{l=2}^{\infty}\frac{(-1)^{l}}{l!}\partial_{I_{l-2}}\bigg(\frac{\partial_{t}^{2}\hat{M}_{ijI_{l-2}}(u)}{r}\bigg)\right)^{TT} -\frac{8G}{c^{4}}\left(\sum_{l=2}^{\infty}\frac{(-1)^{l}l}{(l+1)!}\partial_{aI_{l-2}}\bigg(\frac{\epsilon_{ab(i}\partial_{|t|}\hat{S}_{j)bI_{l-2}}(u)}{r}\bigg)\right)^{TT},
\end{array}\right.
\end{equation}
where the symbol $TT$ represents the TT projection operator, which is defined, for an arbitrary symmetric spatial tensor $X_{ij}$, as
\begin{equation}
\label{equ3.22}(X_{ij})^{TT}:=P_{ik}P_{jl}X_{kl}-\frac{1}{2}P_{ij}P_{kl}X_{kl},
\end{equation}
where
\begin{equation}
\label{equ3.23}P_{ij}=\delta_{ij}-n_{i}n_{j}.
\end{equation}

The effective stress-energy of GWs under the de Donder condition in linearized $f(R)$ gravity is
\begin{align}
\label{equ3.24}t^{\mu\nu}_{f}=\Big<(\tau^{\mu\nu}_{f})^{(2)}_{\rm{rad}}\Big>=t^{\mu\nu}_{GR}\big(\tilde h^{\alpha\beta}\big)+\frac{6a^{2}}{\kappa}\Big<\partial^{\mu}R^{(1)}\partial^{\nu}R^{(1)}\Big>,
\end{align}
where $\big<\cdots\big>$ is the the average over a small spatial volume (several wavelengths)
surrounding each point~\cite{Berry:2011pb}, and
\begin{equation}
\label{equ3.25}t^{\mu\nu}_{GR}(\tilde h^{\alpha\beta})=\frac{1}{4\kappa}\Big<\partial^{\mu}\tilde{h}_{\alpha\beta}\partial^{\nu}\tilde{h}^{\alpha\beta}
-\frac{1}{2}\partial^{\mu}\tilde{h}\partial^{\nu}\tilde{h}\Big>
\end{equation}
is the effective stress-energy of GWs under the de Donder condition in linearized GR with the replace of the variables $h^{\alpha\beta}$ by $\tilde h^{\alpha\beta}$~\cite{Carroll2004}.  The relevant rules for the average are
\begin{align}
\label{equ3.26}\big<\partial_{\mu}X\big>=0,\quad
\big<A(\partial_{\mu}B)\big>=-\big<(\partial_{\mu}A)B\big>,
\end{align}
where $X,A,B$ are three arbitrary quantities.
According to the Refs.~\cite{Carroll2004,Maggiore2008}, $t^{\mu\nu}_{GR}(\tilde h^{\alpha\beta})$ can also be simplified by imposing the TT gauge outside the source, namely,
\begin{equation}
\label{equ3.27}t^{\mu\nu}_{GR}\big(\tilde h^{\alpha\beta}\big)=\frac{1}{4\kappa}\Big<\partial^{\mu}\tilde{h}_{pq}^{TT}\partial^{\nu}\tilde{h}_{pq}^{TT}\Big>.
\end{equation}

The scalar part $R^{(1)}$ satisfies the massive Klein-Gordon equation with an external source.  Its multipole expansion under the condition $r'/r\lll 1$
derived in the spirit of Ref.~\cite{Campbell:1977jf} is
\begin{align}
\label{equ3.28}R^{(1)}(t,\boldsymbol{x})
&=-\frac{m^{2}\kappa}{4\pi}\sum_{l=0}^{\infty}\frac{(-1)^l}{l!}\int d^{3}x'\frac{c}{r'}\hat{X}'_{I_{l}}(\theta',\varphi')\hat{T}_{I_{l}}(t,\boldsymbol{x},\boldsymbol{x}'),
\end{align}
where $r'=|\boldsymbol{x}'|$,
\begin{align}
\label{equ3.29}
X'_{I_{l}}(\theta',\varphi')=&X'_{i_{1}i_{2}\cdots i_{l}}(\theta',\varphi'):= x'_{i_{1}}x'_{i_{2}}\cdots x'_{i_{l}}={r'}^l N_{I_l}(\theta',\varphi'),\\
\label{equ3.30}\hat{T}_{I_{l}}(t,\boldsymbol{x},\boldsymbol{x}')=&\frac{(2l+1)!!}{2^{l+1}l!}\left(\int_{u-\frac{r'}{c}}^{u+\frac{r'}{c}}dt'\left[\hat{\partial}_{I_{l}}\left(\frac{1}{r}\Big(1-\frac{c^2}{r'^2}(t'-u)^2\Big)^l\right)\right]
T^{(1)}(t',\boldsymbol{x}')\right.\notag\\
&\left.-\int_{-\infty}^{u-\frac{r'}{c}}dt'\left[\hat{A}(t,r;t',r')\hat{\partial}_{I_{l}}\left(\frac{1}{r}\Big(1-\frac{c^2}{r'^2}(t'-u)^2\Big)^l\right)\right]
T^{(1)}(t',\boldsymbol{x}')\right)
\end{align}
with $r=|\boldsymbol{x}|$,
\begin{align}
\label{equ3.31}\hat{A}(t,r;t',r'):&=\exp{\Big(\frac{m^2rr'}{2}(\text{int}\ t')_{-}\Big)}-\exp{\Big(\frac{m^2rr'}{2}(\text{int}\ t')_{+}\Big)},\\
\label{equ3.32}(\text{int}\ t')_{\pm}\tilde{f}(t'):&=-\frac{c}{r'}\int_{t'}^{u\pm\frac{r'}{c}}\tilde{f}(\tau)\Big(1+\frac{c}{r}(u-\tau)\Big)d\tau.
\end{align}
Above results show that there exist monopole and dipole radiations for $f(R)$ gravity, which makes its prediction about GWs different from the case in GR.

Now, the $1/r$-expansion in the distance to the source can be applied to Eqs.~\eqref{equ3.21} to obtain
the multipole expansion of $\tilde{h}^{\mu\nu}$ in the radiation field. By Eq.~\eqref{equ2.15}, we have
\begin{align}
\label{equ3.33}&\hat{\partial}_{I_{l}}\left(\frac{F(u)}{r}\right)=\hat{N}_{I_{l}}(\theta,\varphi)\frac{(-1)^{l}}{c^{l}}
\frac{\partial_{u}^{l}F(u)}{r}+O\left(\frac{1}{r^2}\right),
\end{align}
where $O(1/r^2)$ contains the terms whose orders are equal to or higher than $1/r^2$.
Then, we obtain the multipole expansions of $\tilde{h}^{\mu\nu}$ in the radiation field:
\begin{equation}\label{equ3.34}
\left\{\begin{array}{l}
\displaystyle\tilde{h}^{00}_{TT}(t,\boldsymbol{x})= 0,\bigskip\\
\displaystyle\tilde{h}^{0i}_{TT}(t,\boldsymbol{x})= 0,\medskip\\
\displaystyle\tilde{h}^{ij}_{TT}(t,\boldsymbol{x})=-\frac{4G}{c^{2}r}\sum_{l=2}^{\infty}\left(\frac{1}{c^{l}l!}
\Big(\partial_{u}^{l}\hat{M}_{ijI_{l-2}}(u)N_{I_{l-2}}(\theta,\varphi)\Big)^{TT} -\frac{2l}{c^{l+1}(l+1)!}\Big(\epsilon^{}_{ab(i}\partial_{|u|}^{l}\hat{S}_{j)bI_{l-2}}(u)N_{aI_{l-2}}(\theta,\varphi)\Big)^{TT}\right)\\
\displaystyle\qquad\qquad\qquad+O\left(\frac{1}{r^2}\right),
\end{array}\right.
\end{equation}
which is the same as that of $h^{\mu\nu}$ in linearized GR~\cite{Thorne:1980ru,Blanchet:2013haa}.

Similarly to the tensor part, the $1/r$-expansion in the distance to the source should be applied to  Eqs.~(\ref{equ3.28}) and (\ref{equ3.30}) to obtain the multipole expansion of $R^{(1)}$ in the radiation field. They are
\begin{align}
\label{equ3.35}R^{(1)}(t,\boldsymbol{x})&=-\frac{m^{2}\kappa}{4\pi r}\sum_{l=0}^{\infty}\frac{1}{c^{l}l!}\hat{K}_{I_{l}}(t,r)N_{I_{l}}(\theta,\varphi)+O\left(\frac{1}{r^2}\right),
\end{align}
where the effective $l$-pole radiation moment is
\begin{align}
\label{equ3.36}\hat{K}_{I_{l}}(t,r)=&\frac{(2l+1)!!}{2^{l+1}l!}\int d^{3}x'\frac{c}{r'}\hat{X'}_{I_{l}}(\theta',\varphi')\left(\int_{u-\frac{r'}{c}}^{u+\frac{r'}{c}}dt'\bigg[\frac{d^l}{du^l}\Big(1-\frac{c^2}{r'^2}(t'-u)^2\Big)^l\bigg]
T^{(1)}(t',\boldsymbol{x}')\right.\notag\\
&-\left.\int_{-\infty}^{u-\frac{r'}{c}}dt'\left[\hat{A}(t,r;t',r')\frac{d^l}{du^l}\Big(1-\frac{c^2}{r'^2}(t'-u)^2\Big)^l\right]T^{(1)}(t',\boldsymbol{x}')\right).
\end{align}

Upon obtaining (\ref{equ3.34}) and (\ref{equ3.35}), we can write down the multipole expansion of $h^{\mu\nu}$ under the TT gauge in the radiation field by (\ref{equ3.15}). It is
\begin{equation}\label{equ3.37}
\left\{\begin{array}{l}
\displaystyle h^{00}(t,\boldsymbol{x})=-\frac{2G}{3c^4 r}\sum_{l=0}^{\infty}\frac{1}{c^{l}l!}\hat{K}_{I_{l}}(t,r)N_{I_{l}}(\theta,\varphi)+O\left(\frac{1}{r^2}\right),\smallskip\\
\displaystyle h^{0i}(t,\boldsymbol{x})= 0,\medskip\\
\displaystyle h^{ij}(t,\boldsymbol{x})=-\frac{4G}{c^{2}r}\sum_{l=2}^{\infty}\left(\frac{1}{c^{l}l!}\Big(\partial_{u}^{l}\hat{M}_{ijI_{l-2}}(u)N_{I_{l-2}}(\theta,\varphi)\Big)^{TT} -\frac{2l}{c^{l+1}(l+1)!}\Big(\epsilon^{}_{ab(i}\partial_{|u|}^{l}\hat{S}_{j)bI_{l-2}}(u)N_{aI_{l-2}}(\theta,\varphi)\Big)^{TT}\right)\medskip\\
\displaystyle \phantom{h^{ij}(t,\boldsymbol{x})=}+\frac{2G}{3c^4 r}\delta^{ij}\sum_{l=0}^{\infty}\frac{1}{c^{l}l!}\hat{K}_{I_{l}}(t,r)N_{I_{l}}(\theta,\varphi)+O\left(\frac{1}{r^2}\right).
\end{array}\right.
\end{equation}


\section{Energy, momentum, and angular momentum in the GWs for linearized $f(R)$ gravity~\label{Sec:EneggyMomentumAgular}}
Having obtained the multipole expansion of linearized $f(R)$ gravity in the radiation field under the TT gauge, we can deal with its energy, momentum, and angular momentum in the GWs. For the linearized GR, the
fluxes of the energy and the momentum can be evaluated by using the effective stress-energy of GWs,
but the flux of the angular momentum cannot~\cite{Thorne:1980ru}. Similar problem also appears in the linearized $f(R)$ gravity. Now, we, following Refs.~\cite{Thorne:1980ru,Peters:1964zz}, calculate the energy, momentum, and angular momentum in the GWs for linearized $f(R)$ gravity in a unified way.

For $f(R)$ gravity, the energy of the gravitational perturbation inside a volume $V$ enclosed by a large sphere $S$ is
\begin{equation}
\label{equ4.1}E=\int_{V}d^{3}x\ \tau^{00}_{f},
\end{equation}
and the momentum and the angular momentum of the gravitational perturbation in the volume $V$ are
\begin{align}
\label{equ4.2}P_{a}&=\frac{1}{c}\int_{V}d^{3}x\ \tau^{0a}_{f},\\
\label{equ4.3}J_{a}&=\frac{1}{c}\int_{V}d^{3}x\ \epsilon_{abc}x_{b}\tau^{0c}_{f},
\end{align}
respectively. By Eq.~(\ref{equ3.9}) and the Gauss theorem,
\begin{align}
\label{equ4.4}\frac{dE}{dt}&=c\int_{V}d^{3}x\,\partial_{0}\tau^{00}_{f}=-c\int_{V}d^{3}x\,\partial_{i}\tau^{i0}_{f}=-c\int
d\Omega\,r^2n_{i}\tau^{i0}_{f},\\
\label{equ4.5}\frac{dP_{a}}{dt}&=\int_{V}d^{3}x\,\partial_{0}\tau^{0a}_{f}=-\int_{V}d^{3}x\,\partial_{i}\tau^{ia}_{f}=-\int d\Omega\,r^2n_{i}\tau^{ia}_{f},\\
\label{equ4.6}\frac{dJ_{a}}{dt}&=\int_{V}d^{3}x\,\partial_{0}(\epsilon_{abc}x_{b}\tau^{0c}_{f})
=\int_{V}d^{3}x\,\epsilon_{abc}x_{b}\partial_{0}\tau^{0c}_{f}
=-\int_{V}d^{3}x\,\epsilon_{abc}x_{b}\partial_{i}\tau^{ic}_{f}\notag\\
&=-\int_{V}d^{3}x\,\partial_{i}(\epsilon_{abc}x_{b}\tau^{ic}_{f})
=-\int d\Omega\,r^3\epsilon_{abc}n_{i}n_{b}\tau^{ic}_{f},
\end{align}
where $n_{i}$ is also the unit normal vector of the large sphere $S$, $d\Omega$ is the element of solid angle and its integral domain is $4\pi$. Thus, the fluxes of energy, momentum, and angular momentum carried by the outward-propagating GW should be~\cite{Thorne:1980ru}
\begin{align}
\label{equ4.7}\frac{dE}{d\Omega dt}&=+cr^2\Big<n_{i}(\tau^{i0}_{f})^{(2)}_{\rm{rad}}\Big>,\\
\label{equ4.8}\frac{dP_{a}}{d\Omega dt}&=+r^2\Big<n_{i}(\tau^{ia}_{f})^{(2)}_{\rm{rad}}\Big>,\\
\label{equ4.9}\frac{dJ_{a}}{d\Omega dt}&=+r^3\Big<\epsilon_{abc}n_{i}n_{b}(\tau^{ic}_{f})^{(2)}_{\rm{rad}}\Big>,
\end{align}
respectively, where $(\tau^{\mu\nu}_{f})^{(2)}_{\rm{rad}}$ is the quadratic term of the stress-energy pseudotensor under the TT gauge in the radiation field, and the higher-order terms of $\tau^{\mu\nu}_{f}$ have been omitted, because only the linearized $f(R)$ gravity is discussed~\cite{Peters:1964zz}. Next, we are to derive
\begin{align}
\label{equ4.10}\frac{dE}{d\Omega dt}&=\left(\frac{dE}{d\Omega dt}\right)^{[0]}+O\left(\frac{1}{r}\right),\\
\label{equ4.11}\frac{dP_{a}}{d\Omega dt}&=\left(\frac{dP_{a}}{d\Omega dt}\right)^{[0]}+O\left(\frac{1}{r}\right),\\
\label{equ4.12}\frac{dJ_{a}}{d\Omega dt}&=\left(\frac{dJ_{a}}{d\Omega dt}\right)^{[0]}+O\left(\frac{1}{r}\right),
\end{align}
where
\begin{align}
\label{equ4.13}\left(\frac{dE}{d\Omega dt}\right)^{[0]}&=cr^2\Big<n_{i}(\tau^{i0}_{f})^{(2)}_{\rm{rad}}\Big>^{[2]},\\
\label{equ4.14}\left(\frac{dP_{a}}{d\Omega dt}\right)^{[0]}&=r^2\Big<n_{i}(\tau^{ia}_{f})^{(2)}_{\rm{rad}}\Big>^{[2]},\\
\label{equ4.15}\left(\frac{dJ_{a}}{d\Omega dt}\right)^{[0]}&=r^3\Big<\epsilon_{abc}n_{i}n_{b}(\tau^{ic}_{f})^{(2)}_{\rm{rad}}\Big>^{[3]},
\end{align}
are the leading terms of the fluxes of energy, momentum, and angular momentum in the GWs for linearized $f(R)$ gravity, respectively, and the superscript $[m]\ (m=0,1,2,3\cdots)$ represents that the term of the $1/r^m$-order for the corresponding quantity is taken. In the following, the fluxes of the energy, momentum, and the angular momentum
are used to denote their leading terms, respectively. It should be noted that if $\Big<\epsilon_{abc}n_{i}n_{b}(\tau^{ic}_{f})^{(2)}_{\rm{rad}}\Big>^{[2]}$  did not vanish, (\ref{equ4.9}) would result in the divergence of the flux of angular momentum carried by the outward-propagating GW. So, we also need to prove that it is impossible. In other words,
\begin{align}
\label{equ4.16}\Big<\epsilon_{abc}n_{i}n_{b}(\tau^{ic}_{f})^{(2)}_{\rm{rad}}\Big>^{[2]}=0.
\end{align}

By (\ref{equ3.12}), $(\tau^{i0}_{f})^{(2)}_{\rm{rad}}$ and $(\tau^{ij}_{f})^{(2)}_{\rm{rad}}$ under the TT gauge are
\begin{align}
\label{equ4.17}(\tau^{i0}_{f})^{(2)}_{\rm{rad}}=&\frac{1}{2\kappa}\bigg(-\frac{1}{2}\partial_{0}\tilde{h}_{pq}^{TT}\partial_{i}\tilde{h}_{pq}^{TT}
+\partial_{0}\tilde{h}_{pq}^{TT}\partial_{p}\tilde{h}_{iq}^{TT}-12a^{2}\partial_{0}R^{(1)}\partial_{i}R^{(1)}\bigg),\\
\label{equ4.18}(\tau^{ij}_{f})^{(2)}_{\rm{rad}}=&\frac{1}{2\kappa}\bigg(-\tilde{h}_{pq}^{TT}\partial_{p}\partial_{q}\tilde{h}_{ij}^{TT}
+\frac{1}{2}\partial_{i}\tilde{h}_{pq}^{TT}\partial_{j}\tilde{h}_{pq}^{TT}
-\partial_{0}\tilde{h}_{ip}^{TT}\partial_{0}\tilde{h}_{jp}^{TT}+\partial_{q}\tilde{h}_{ip}^{TT}\partial_{q}\tilde{h}_{jp}^{TT}+\partial_{p}\tilde{h}_{iq}^{TT}\partial_{q}\tilde{h}_{jp}^{TT}
-\partial_{i}\tilde{h}_{pq}^{TT}\partial_{p}\tilde{h}_{jq}^{TT}\notag\\
&-\partial_{j}\tilde{h}_{pq}^{TT}\partial_{p}\tilde{h}_{iq}^{TT}+\delta_{ij}\Big(\frac{1}{4}\partial_{0}\tilde{h}_{pq}^{TT}\partial_{0}\tilde{h}_{pq}^{TT}-\frac{1}{4}\partial_{m}\tilde{h}_{pq}^{TT}\partial_{m}\tilde{h}_{pq}^{TT}
+\frac{1}{2}\partial_{p}\tilde{h}_{qm}^{TT}\partial_{q}\tilde{h}_{pm}^{TT}\Big)
+12a^{2}\partial_{i}R^{(1)}\partial_{j}R^{(1)}\notag\\
&-\delta_{ij}\Big(a{\Big(R^{(1)}\Big)}^2+6a^2\partial^{\alpha}R^{(1)}\partial_{\alpha}R^{(1)}\Big)\bigg).
\end{align}
Then, one should calculate the average part of Eqs. \eqref{equ4.13}---\eqref{equ4.16}.
Firstly, the average part in \eqref{equ4.13} is
\begin{align}
\label{equ4.19}\Big<n_{i}(\tau^{i0}_{f})^{(2)}_{\rm{rad}}\Big>^{[2]}=&\frac{1}{2\kappa}\left(
-\frac{1}{2}\Big<n_{i}\Big(\partial_{0}\tilde{h}_{pq}^{TT}\Big)^{[1]}\Big(\partial_{i}\tilde{h}_{pq}^{TT}\Big)^{[1]}\Big>
+\Big<n_{i}\Big(\partial_{0}\tilde{h}_{pq}^{TT}\Big)^{[1]}\Big(\partial_{p}\tilde{h}_{iq}^{TT}\Big)^{[1]}\Big>
-12a^{2}\Big<n_{i}\partial_{0}R^{(1)}\partial_{i}R^{(1)}\Big>^{[2]}\right).
\end{align}
From (\ref{equ3.34}),
\begin{equation}\label{equ4.20}
\Big(\tilde{h}_{pq}^{TT}\Big)^{[1]}=-\frac{4G}{c^{2}r}\sum_{l=2}^{\infty}\left(\frac{1}{c^{l}l!}\Big(\partial_{u}^{l}\hat{M}_{pqI_{l-2}}(u)N_{I_{l-2}}(\theta,\varphi)\Big)^{TT} -\frac{2l}{c^{l+1}(l+1)!}\Big(\epsilon^{}_{ab(p}\partial_{|u|}^{l}\hat{S}_{q)bI_{l-2}}(u)N_{aI_{l-2}}(\theta,\varphi)\Big)^{TT}\right),
\end{equation}
and then, the combination of (\ref{equ2.13}), (\ref{equ2.14}), (\ref{equ3.22}), (\ref{equ3.23}), and (\ref{equ4.20}) brings about~\cite{Peters:1964zz}
\begin{align}
\label{equ4.21}\Big(\partial_{k}\tilde{h}_{pq}^{TT}\Big)^{[1]}&=\Big(\partial_{k}\Big(\tilde{h}_{pq}^{TT}\Big)^{[1]}\Big)^{[1]}\notag\\
&=n_{k}\frac{4G}{c^{3}r}\sum_{l=2}^{\infty}\left(\frac{1}{c^{l}l!}\Big(\partial_{u}^{l+1}\hat{M}_{pqI_{l-2}}(u)N_{I_{l-2}}(\theta,\varphi)\Big)^{TT} -\frac{2l}{c^{l+1}(l+1)!}\Big(\epsilon_{ab(p}\partial_{|u|}^{l+1}\hat{S}_{q)bI_{l-2}}(u)N_{aI_{l-2}}(\theta,\varphi)\Big)^{TT}\right)\notag\\
&=-n_{k}\Big(\partial_{0}\tilde{h}_{pq}^{TT}\Big)^{[1]}=-n_{k}\partial_{0}\Big(\tilde{h}_{pq}^{TT}\Big)^{[1]}.
\end{align}
Thus, \eqref{equ4.19} results in
\begin{align}
\label{equ4.22}\Big<n_{i}(\tau^{i0}_{f})^{(2)}_{\rm{rad}}\Big>^{[2]}
=&\frac{1}{4\kappa}
\Big<\partial_{0}\Big(\tilde{h}_{pq}^{TT}\Big)^{[1]}\partial_{0}\Big(\tilde{h}_{pq}^{TT}\Big)^{[1]}\Big>
-\frac{6a^{2}}{\kappa}\Big<\partial_{0}R^{(1)}\partial_{r}R^{(1)}\Big>^{[2]},
\end{align}
where \eqref{equ2.2}, \eqref{equ4.21}, and the de Donder condition under the TT gauge,
\begin{align}
\label{equ4.23}\partial_{p}\tilde{h}_{pq}^{TT}=0
\end{align}
have been used.  Substituting \eqref{equ4.22} in \eqref{equ4.13} immediately gives
\begin{align}
\label{equ4.24}\left(\frac{dE}{d\Omega dt}\right)^{[0]}&=\frac{c r^2}{4\kappa}\Big<\partial_{0}\tilde{h}_{pq}^{TT}\partial_{0}\tilde{h}_{pq}^{TT}\Big>^{[2]}-\frac{6cr^2a^2}{\kappa}\Big<\partial_{r}R^{(1)}\partial_{0}R^{(1)}\Big>^{[2]}.
\end{align}
Secondly,
\begin{align}
\label{equ4.25}\Big<n_{i}(\tau^{ia}_{f})^{(2)}_{\rm{rad}}\Big>^{[2]}=&\frac{1}{2\kappa}\bigg(
-\Big<n_{i}\Big(\tilde{h}_{pq}^{TT}\Big)^{[1]}\Big(\partial_{p}\partial_{q}\tilde{h}_{ia}^{TT}\Big)^{[1]}\Big>
+\frac{1}{2}\Big<n_{i}\Big(\partial_{i}\tilde{h}_{pq}^{TT}\Big)^{[1]}\Big(\partial_{a}\tilde{h}_{pq}^{TT}\Big)^{[1]}\Big>
-\Big<n_{i}\partial_{0}\tilde{h}_{ip}^{TT}\partial_{0}\tilde{h}_{ap}^{TT}\Big>^{[2]}\notag\\
&+\Big<n_{i}\Big(\partial_{q}\tilde{h}_{ip}^{TT}\Big)^{[1]}\Big(\partial_{q}\tilde{h}_{ap}^{TT}\Big)^{[1]}\Big>
+\Big<n_{i}\Big(\partial_{p}\tilde{h}_{iq}^{TT}\Big)^{[1]}\Big(\partial_{q}\tilde{h}_{ap}^{TT}\Big)^{[1]}\Big>
-\Big<n_{i}\Big(\partial_{i}\tilde{h}_{pq}^{TT}\Big)^{[1]}\Big(\partial_{p}\tilde{h}_{aq}^{TT}\Big)^{[1]}\Big>\notag\\
&-\Big<n_{i}\Big(\partial_{a}\tilde{h}_{pq}^{TT}\Big)^{[1]}\Big(\partial_{p}\tilde{h}_{iq}^{TT}\Big)^{[1]}\Big>
+\frac{1}{4}\Big<n_{a}\partial_{0}\tilde{h}_{pq}^{TT}\partial_{0}\tilde{h}_{pq}^{TT}\Big>^{[2]}
-\frac{1}{4}\Big<n_{a}\Big(\partial_{m}\tilde{h}_{pq}^{TT}\Big)^{[1]}\Big(\partial_{m}\tilde{h}_{pq}^{TT}\Big)^{[1]}\Big>\notag\\
&+\frac{1}{2}\Big<n_{a}\Big(\partial_{p}\tilde{h}_{qm}^{TT}\Big)^{[1]}\Big(\partial_{q}\tilde{h}_{pm}^{TT}\Big)^{[1]}\Big>
+12a^{2}\Big<n_{i}\partial_{i}R^{(1)}\partial_{a}R^{(1)}\Big>^{[2]}
-a\Big<n_{a}{\Big(R^{(1)}\Big)}^2\Big>^{[2]}\notag\\
&-6a^2\Big<n_{a}\partial^{\alpha}R^{(1)}\partial_{\alpha}R^{(1)}\Big>^{[2]}\bigg) \notag \\
=&\frac{1}{2\kappa}\bigg(-\frac{1}{2}\Big<\partial_{0}\tilde{h}_{pq}^{TT}\partial_{a}\tilde{h}_{pq}^{TT}\Big>^{[2]}
+12a^{2}\Big<\partial_{r}R^{(1)}\partial_{a}R^{(1)}\Big>^{[2]}-a\Big<n_{a}{\Big(R^{(1)}\Big)}^2\Big>^{[2]}
-6a^2\Big<n_{a}\partial^{\alpha}R^{(1)}\partial_{\alpha}R^{(1)}\Big>^{[2]}\bigg) \notag \\
=&-\frac{1}{4\kappa}\Big<\partial_{0}\tilde{h}_{pq}^{TT}\partial_{a}\tilde{h}_{pq}^{TT}\Big>^{[2]}
+\frac{6a^{2}}{\kappa}\Big<\partial_{r}R^{(1)}\partial_{a}R^{(1)}\Big>^{[2]}.
\end{align}
In the second step, \eqref{equ2.2}, \eqref{equ3.26}, \eqref{equ4.21}, \eqref{equ4.23}, and the following formula are used,
\begin{align}
\label{equ4.26}\Big(\partial_{j}\partial_{k}\tilde{h}_{pq}^{TT}\Big)^{[1]}&=\Big(\partial_{j}\Big(\partial_{k}\tilde{h}_{pq}^{TT}\Big)^{[1]}\Big)^{[1]}
=-n_{j}\partial_{0}\Big(\partial_{k}\tilde{h}_{pq}^{TT}\Big)^{[1]}=n_{j}n_{k}\Big(\partial_{0}^{2}\tilde{h}_{pq}^{TT}\Big)^{[1]}
=n_{j}n_{k}\partial_{0}^{2}\Big(\tilde{h}_{pq}^{TT}\Big)^{[1]},
\end{align}
where its proof is the same as that of \eqref{equ4.21}.
The last step is valid because of \eqref{equ2.13}, \eqref{equ3.26}, and the equation of $R^{(1)}$ in the radiation field,
\begin{align}
\label{equ4.27}\square_{\eta} R^{(1)}=\frac{1}{6a}R^{(1)}.
\end{align}
Combined with \eqref{equ4.14}, Eq.~\eqref{equ4.25} gives
\begin{align}
\label{equ4.28}\left(\frac{dP_{a}}{d\Omega dt}\right)^{[0]}&=-\frac{ r^2}{4\kappa}\Big<\partial_{0}\tilde{h}_{pq}^{TT}\partial_{a}\tilde{h}_{pq}^{TT}\Big>^{[2]}+\frac{6r^2a^2}{\kappa}\Big<\partial_{r}R^{(1)}\partial_{a}R^{(1)}\Big>^{[2]}.
\end{align}
Thirdly, we need to calculate the average part of (\ref{equ4.15}). Since the calculation is lengthy,
we leave it in Appendix and just give the result here
\begin{align}
\label{equ4.29}\Big<\epsilon_{abc}n_{i}n_{b}(\tau^{ic}_{f})^{(2)}_{\rm{rad}}\Big>^{[3]}
=&\frac{1}{2r\kappa}\left(\Big<\epsilon_{abc}\Big(\tilde{h}_{bi}^{TT}\partial_{0}\tilde{h}_{ic}^{TT}
-\frac{1}{2}x_{b}\partial_{c}\tilde{h}_{pq}^{TT}\partial_{0}\tilde{h}_{pq}^{TT}\Big)\Big>^{[3]}
+\frac{6a^{2}}{\kappa}\Big<\epsilon_{abc}n_{b}\partial_{r}R^{(1)}\partial_{c}R^{(1)}\Big>^{[3]}\right).
\end{align}
Therefore, from \eqref{equ4.15},
\begin{align}
\label{equ4.30}\left(\frac{dJ_{a}}{d\Omega dt}\right)^{[0]}&=\frac{r^2}{2\kappa}\Big<\epsilon_{abc}\Big(\tilde{h}_{bi}^{TT}\partial_{0}\tilde{h}_{ic}^{TT}-\frac{1}{2}x_{b}\partial_{c}\tilde{h}_{pq}^{TT}\partial_{0}\tilde{h}_{pq}^{TT}\Big)\Big>^{[2]}
+\frac{6r^3a^2}{\kappa}\Big<\epsilon_{abc}n_{b}\partial_{r}R^{(1)}\partial_{c}R^{(1)}\Big>^{[3]}.
\end{align}
Finally,
\begin{align}
\label{equ4.31}\Big<\epsilon_{abc}n_{i}n_{b}(\tau^{ic}_{f})^{(2)}_{\rm{rad}}\Big>^{[2]}=&\frac{1}{2\kappa}\epsilon_{abc}\bigg(
-\Big<n_{i}n_{b}\Big(\tilde{h}_{pq}^{TT}\Big)^{[1]}\Big(\partial_{p}\partial_{q}\tilde{h}_{ic}^{TT}\Big)^{[1]}\Big>
+\frac{1}{2}\Big<n_{i}n_{b}\Big(\partial_{i}\tilde{h}_{pq}^{TT}\Big)^{[1]}\Big(\partial_{c}\tilde{h}_{pq}^{TT}\Big)^{[1]}\Big>
\notag\\
&-\Big<n_{i}n_{b}\partial_{0}\tilde{h}_{ip}^{TT}\partial_{0}\tilde{h}_{cp}^{TT}\Big>^{[2]}
+\Big<n_{i}n_{b}\Big(\partial_{q}\tilde{h}_{ip}^{TT}\Big)^{[1]}\Big(\partial_{q}\tilde{h}_{cp}^{TT}\Big)^{[1]}\Big>
+\Big<n_{i}n_{b}\Big(\partial_{p}\tilde{h}_{iq}^{TT}\Big)^{[1]}\Big(\partial_{q}\tilde{h}_{cp}^{TT}\Big)^{[1]}\Big>\notag\\
&-\Big<n_{i}n_{b}\Big(\partial_{i}\tilde{h}_{pq}^{TT}\Big)^{[1]}\Big(\partial_{p}\tilde{h}_{cq}^{TT}\Big)^{[1]}\Big>
-\Big<n_{i}n_{b}\Big(\partial_{c}\tilde{h}_{pq}^{TT}\Big)^{[1]}\Big(\partial_{p}\tilde{h}_{iq}^{TT}\Big)^{[1]}\Big>\notag\\
&+12a^{2}\Big<n_{i}n_{b}\Big(\partial_{i}R^{(1)}\Big)^{[1]}\Big(\partial_{c}R^{(1)}\Big)^{[1]}\Big>\bigg) \notag\\
=&\frac{6a^{2}}{\kappa}\epsilon_{abc}\Big<n_{i}n_{b}\Big(\partial_{i}R^{(1)}\Big)^{[1]}\Big(\partial_{c}R^{(1)}\Big)^{[1]}\Big>.
\end{align}
In this derivation, \eqref{equ3.26}, \eqref{equ4.21}, \eqref{equ4.23}, \eqref{equ4.26}, and $\epsilon_{abc}n_bn_c=0$ have been used. By (\ref{equ2.13}), \eqref{equ3.35} brings about
\begin{align}
\label{equ4.32}\Big(\partial_{k}R^{(1)}\Big)^{[1]}&=\Big(\partial_{k}\Big(R^{(1)}\Big)^{[1]}\Big)^{[1]}=n_{k}\Big(\partial_{r}\Big(R^{(1)}\Big)^{[1]}\Big)^{[1]},
\end{align}
so the last term in \eqref{equ4.31} vanishes due to $\epsilon_{abc}n_bn_c=0$ again.  Hence, \eqref{equ4.16} holds, and it ensures that the flux of angular momentum carried by the outward-propagating GW does not diverge.

As mentioned before, the fluxes of energy and momentum carried by the outward-propagating GW in linearized $f(R)$ gravity could also be directly expressed in terms of the effective stress-energy $t^{\mu\nu}_{f}$, namely,
\begin{align}
\label{equ4.33}\frac{dE}{d\Omega dt}&=+c r^2n_{i}t^{i0}_{f}=-\frac{c r^2}{4\kappa}n_{i}\Big<\partial_{i}\tilde{h}_{pq}^{TT}\partial_{0}\tilde{h}_{pq}^{TT}\Big>-\frac{6cr^2a^2}{\kappa}n_{i}\Big<\partial_{i}R^{(1)}\partial_{0}R^{(1)}\Big>,\\
\label{equ4.34}\frac{dP_{a}}{d\Omega dt}&=+r^2n_{i}t^{ia}_{f}=\frac{ r^2}{4\kappa}n_{i}\Big<\partial_{i}\tilde{h}_{pq}^{TT}\partial_{a}\tilde{h}_{pq}^{TT}\Big>+\frac{6r^2a^2}{\kappa}n_{i}\Big<\partial_{i}R^{(1)}\partial_{a}R^{(1)}\Big>,
\end{align}
where Eqs.~(\ref{equ3.24}) and (\ref{equ3.27}) have been used. Because $n_{i}$ are not only the unit normal vector of the large sphere $S$ but also the components of the unit radial vector, following the method in Refs.~\cite{Carroll2004,Maggiore2008}, we have
\begin{align}
\label{equ4.35}\frac{dE}{d\Omega dt}&=-\frac{c r^2}{4\kappa}\Big<\partial_{r}\tilde{h}_{pq}^{TT}\partial_{0}\tilde{h}_{pq}^{TT}\Big>-\frac{6cr^2a^2}{\kappa}\Big<\partial_{r}R^{(1)}\partial_{0}R^{(1)}\Big>,\\
\label{equ4.36}\frac{dP_{a}}{d\Omega dt}&=\frac{ r^2}{4\kappa}\Big<\partial_{r}\tilde{h}_{pq}^{TT}\partial_{a}\tilde{h}_{pq}^{TT}\Big>+\frac{6r^2a^2}{\kappa}\Big<\partial_{r}R^{(1)}\partial_{a}R^{(1)}\Big>.
\end{align}
Thus, (\ref{equ4.35}) and (\ref{equ4.36}) can recover (\ref{equ4.24}) and (\ref{equ4.28}), respectively, with the help of
\begin{align}
\label{equ4.37}\Big(\partial_{r}\tilde{h}_{pq}^{TT}\Big)^{[1]}&=\Big(n_{k}\partial_{k}\tilde{h}_{pq}^{TT}\Big)^{[1]}=n_{k}\Big(\partial_{k}\tilde{h}_{pq}^{TT}\Big)^{[1]}=-\Big(\partial_{0}\tilde{h}_{pq}^{TT}\Big)^{[1]},
\end{align}
where \eqref{equ2.2} and \eqref{equ4.21} have been used. The above fluxes of energy, momentum, and angular momentum in the GWs for linearized $f(R)$ gravity, namely, Eqs. (\ref{equ4.24}), (\ref{equ4.28}), and (\ref{equ4.30}), are reasonable, because they can reduce to the corresponding fluxes in linearized GR~\cite{Thorne:1980ru,Maggiore2008}, respectively,
when $f(R)$ gravity reduces to GR, namely, $f(R)=R$, $f_{R}=1$, $a=0$, and
$\tilde{h}^{\mu\nu}=h^{\mu\nu}$.

Obviously, the fluxes of energy, momentum, and angular momentum in the GWs for linearized $f(R)$ gravity have two parts which are associated with the tensor part and the scalar part in the multipole expansion of linearized $f(R)$ gravity, respectively. The former are the same as those in GR~\cite{Thorne:1980ru,Maggiore2008} with the replace of the variables $h_{pq}^{TT}$ by $\tilde{h}_{pq}^{TT}$, and the latter, as the corrections to the results in GR, result from the massive scalar degree of freedom, $R^{(1)}$, in linearized $f(R)$ gravity. Thus, integrating them over a sphere gives the total power and rates of momentum and angular momentum, respectively,
\begin{align}
\label{equ4.38}\left(\frac{dE}{dt}\right)^{[0]}&=\left(\frac{dE}{dt}\right)^{[0]}_{GR}-\frac{6cr^2a^2}{\kappa}\int
\Big<\partial_{r}R^{(1)}\partial_{0}R^{(1)}\Big>^{[2]}d\Omega,\\
\label{equ4.39}\left(\frac{dP_{a}}{dt}\right)^{[0]}&=\left(\frac{dP_{a}}{dt}\right)^{[0]}_{GR}+\frac{6r^2a^2}{\kappa}
\int\Big<\partial_{r}R^{(1)}\partial_{a}R^{(1)}\Big>^{[2]}d\Omega,\\
\label{equ4.40}\left(\frac{dJ_{a}}{dt}\right)^{[0]}&=\left(\frac{dJ_{a}}{dt}\right)^{[0]}_{GR}+\frac{6r^3a^2}{\kappa}
\int\Big<\epsilon_{abc}n_{b}\partial_{r}R^{(1)}\partial_{c}R^{(1)}\Big>^{[3]}d\Omega
\end{align}
with
\begin{align}
\label{equ4.41}\left(\frac{dE}{dt}\right)^{[0]}_{GR}&=\frac{c r^2}{4\kappa}\int\Big<\partial_{0}\tilde{h}_{pq}^{TT}\partial_{0}\tilde{h}_{pq}^{TT}\Big>^{[2]}d\Omega,\\
\label{equ4.42}\left(\frac{dP_{a}}{dt}\right)^{[0]}_{GR}&=-\frac{ r^2}{4\kappa}\int\Big<\partial_{0}\tilde{h}_{pq}^{TT}\partial_{a}\tilde{h}_{pq}^{TT}\Big>^{[2]}d\Omega,\\
\label{equ4.43}\left(\frac{dJ_{a}}{dt}\right)^{[0]}_{GR}&=\frac{r^2}{2\kappa}\int\Big<\epsilon_{abc}\Big(\tilde{h}_{bi}^{TT}\partial_{0}\tilde{h}_{ic}^{TT}-\frac{1}{2}x_{b}\partial_{c}\tilde{h}_{pq}^{TT}\partial_{0}\tilde{h}_{pq}^{TT}\Big)\Big>^{[2]}d\Omega.
\end{align}
The results of (\ref{equ4.41})---(\ref{equ4.43}) are presented in the Ref.~\cite{Thorne:1980ru}, and they are
\begin{align}
\label{equ4.44}\bigg(\frac{dE}{dt}\bigg)^{[0]}_{GR}=&\sum_{l=2}^{\infty}\frac{G}{c^{2l+1}}\bigg(
\frac{(l+1)(l+2)}{l(l-1)l!(2l+1)!!}\Big<\partial_{u}^{l+1}\hat{M}_{I_{l}}(u)\partial_{u}^{l+1}\hat{M}_{I_{l}}(u)\Big>\notag\\
&\qquad\qquad\qquad\qquad\qquad+\frac{4l(l+2)}{c^2(l-1)(l+1)!(2l+1)!!}\Big<\partial_{u}^{l+1}\hat{S}_{I_{l}}(u)\partial_{u}^{l+1}\hat{S}_{I_{l}}(u)\Big>\bigg),\\
\label{equ4.45}\bigg(\frac{dP_{a}}{dt}\bigg)^{[0]}_{GR}=&\sum_{l=2}^{\infty}\frac{2G}{c^{2l+3}}\bigg(
\frac{(l+2)(l+3)}{l(l+1)!(2l+3)!!}\Big<\partial_{u}^{l+1}\hat{M}_{I_{l}}(u)\partial_{u}^{l+2}\hat{M}_{aI_{l}}(u)\Big>\notag\\
&\qquad\qquad\qquad\qquad\qquad+\frac{4(l+2)}{(l-1)(l+1)!(2l+1)!!}\Big<\epsilon_{abc}\partial_{u}^{l+1}\hat{M}_{bI_{l-1}}(u)\partial_{u}^{l+1}\hat{S}_{cI_{l-1}}(u)\Big>\notag\\
&\qquad\qquad\qquad\qquad\qquad+\frac{4(l+3)}{c^2(l+1)!(2l+3)!!}\Big<\partial_{u}^{l+1}\hat{S}_{I_{l}}(u)\partial_{u}^{l+2}\hat{S}_{aI_{l}}(u)\Big>\bigg),\\
\label{equ4.46}\bigg(\frac{dJ_{a}}{dt}\bigg)^{[0]}_{GR}=&\sum_{l=2}^{\infty}\frac{G}{c^{2l+1}}\bigg(
\frac{(l+1)(l+2)}{(l-1)l!(2l+1)!!}\Big<\epsilon_{abc}\partial_{u}^{l}\hat{M}_{bI_{l-1}}(u)\partial_{u}^{l+1}\hat{M}_{cI_{l-1}}(u)\Big>\notag\\
&\qquad\qquad\qquad\qquad\qquad+\frac{4l^2(l+2)}{c^2(l-1)(l+1)!(2l+1)!!}\Big<\epsilon_{abc}\partial_{u}^{l}\hat{S}_{bI_{l-1}}(u)\partial_{u}^{l+1}\hat{S}_{cI_{l-1}}(u)\Big>\bigg).
\end{align}
The remaining task is to calculate the second terms of (\ref{equ4.38})---(\ref{equ4.40}). Since the integrands in
the second terms of (\ref{equ4.38}) and (\ref{equ4.39}) are the order $1/r^2$ and the integrand in the second term of (\ref{equ4.40}) is the order $1/r^3$, they should be dealt with separately. By (\ref{equ2.13}),
\eqref{equ3.35} gives rise to
\begin{align}
\label{equ4.47}\Big(\partial_{0}R^{(1)}\Big)^{[1]}&=\Big(\partial_{0}\Big(R^{(1)}\Big)^{[1]}\Big)^{[1]}=-\frac{m^{2}\kappa}{4\pi cr}\sum_{l=0}^{\infty}\frac{1}{c^{l}l!}\partial_{t}\hat{K}_{I_{l}}(t,r)N_{I_{l}}(\theta,\varphi),\\
\label{equ4.48}\Big(\partial_{r}R^{(1)}\Big)^{[1]}&=\Big(\partial_{r}\Big(R^{(1)}\Big)^{[1]}\Big)^{[1]}=-\frac{m^{2}\kappa}{4\pi r}\sum_{l=0}^{\infty}\frac{1}{c^{l}l!}\partial_{r}\hat{K}_{I_{l}}(t,r)N_{I_{l}}(\theta,\varphi).
\end{align}
The second terms in (\ref{equ4.38}) and (\ref{equ4.39}) are then
\begin{align}
\label{equ4.49}-\frac{6cr^2a^2}{\kappa}\int\Big<\partial_{r}R^{(1)}&\partial_{0}R^{(1)}\Big>^{[2]}d\Omega
=-\frac{6cr^2a^2}{\kappa}\int\Big<\Big(\partial_{r}R^{(1)}\Big)^{[1]}\Big(\partial_{0}R^{(1)}\Big)^{[1]}\Big>d\Omega\notag\\
&=-\frac{6a^2}{\kappa}\bigg(\frac{m^{2}\kappa}{4\pi}\bigg)^2\sum_{l=0}^{\infty}
\sum_{l'=0}^{\infty}\frac{1}{c^{l+l'}l!l'!}\int\Big<\partial_{r}\hat{K}_{I_{l}}(t,r)N_{I_{l}}(\theta,\varphi)
\partial_{t}\hat{K}_{J_{l'}}(t,r)N_{J_{l'}}(\theta,\varphi)\Big>d\Omega\notag\\
&=-\frac{G}{12\pi c^4}\sum_{l=0}^{\infty}\sum_{l'=0}^{\infty}\frac{1}{c^{l+l'}l!l'!}\Big<\int\partial_{r}\hat{K}_{I_{l}}(t,r)N_{I_{l}}(\theta,\varphi)
\partial_{t}\hat{K}_{J_{l'}}(t,r)N_{J_{l'}}(\theta,\varphi)d\Omega\Big>,\\
\label{equ4.50}+\frac{6r^2a^2}{\kappa}\int\Big<\partial_{r}R^{(1)}&\partial_{a}R^{(1)}\Big>^{[2]}d\Omega
=+\frac{6r^2a^2}{\kappa}\int\Big<\Big(\partial_{r}R^{(1)}\Big)^{[1]}\Big(\partial_{a}R^{(1)}\Big)^{[1]}\Big>d\Omega\notag\\
&=+\frac{6a^2}{\kappa}\bigg(\frac{m^{2}\kappa}{4\pi}\bigg)^2\sum_{l=0}^{\infty}
\sum_{l'=0}^{\infty}\frac{1}{c^{l+l'}l!l'!}\int\Big<\partial_{r}\hat{K}_{I_{l}}(t,r)N_{I_{l}}(\theta,\varphi)
n_{a}\partial_{r}\hat{K}_{J_{l'}}(t,r)N_{J_{l'}}(\theta,\varphi)\Big>d\Omega\notag\\
&=+\frac{G}{12\pi c^4}\sum_{l=0}^{\infty}\sum_{l'=0}^{\infty}\frac{1}{c^{l+l'}l!l'!}\Big<\int\partial_{r}\hat{K}_{I_{l}}(t,r)N_{I_{l}}(\theta,\varphi)
\partial_{r}\hat{K}_{J_{l'}}(t,r)N_{J_{l'}}(\theta,\varphi)n_{a}d\Omega\Big>,
\end{align}
where $m^2=1/6a$ and $\kappa=8\pi G/c^4$ have been used. Further, applying the integral formulas \eqref{equ2.18}
and \eqref{equ2.19} to \eqref{equ4.49} and \eqref{equ4.50} gives, respectively,
\begin{align*}
-\frac{6cr^2a^2}{\kappa}\int\Big<\partial_{r}R^{(1)}\partial_{0}R^{(1)}\Big>^{[2]}d\Omega
&=-\frac{G}{12\pi c^4}\sum_{l=0}^{\infty}\sum_{l'=0}^{\infty}\frac{1}{c^{l+l'}l!l'!}\Big<\partial_{r}\hat{K}_{I_{l}}(t,r)
\partial_{t}\hat{K}_{I_{l'}}(t,r)\Big>\frac{4\pi l!}{(2l+1)!!}\delta_{l\,l'}\notag\\
&=-\sum_{l=0}^{\infty}\frac{G}{3c^{2l+4}}\frac{1}{l!(2l+1)!!}\Big<\partial_{t}\hat{K}_{I_{l}}(t,r)\partial_{r}\hat{K}_{I_{l}}(t,r)\Big>,\\
+\frac{6r^2a^2}{\kappa}\int\Big<\partial_{r}R^{(1)}\partial_{a}R^{(1)}\Big>^{[2]}d\Omega
&=+\frac{G}{12\pi c^4}\sum_{l=0}^{\infty}\sum_{l'=0}^{\infty}\frac{1}{c^{l+l'}l!l'!}\left(\Big<\partial_{r}\hat{K}_{aJ_{l'}}(t,r)
\partial_{r}\hat{K}_{J_{l'}}(t,r)\Big>\frac{4\pi(l'+1)!}{(2l'+3)!!}\delta_{l\,l'+1}\right.\notag\\
&\quad+\left.\Big<\partial_{r}\hat{K}_{I_{l}}(t,r)
\partial_{r}\hat{K}_{aI_{l}}(t,r)\Big>\frac{4\pi(l+1)!}{(2l+3)!!}\delta_{l'\,l+1}\right)\notag\\
&=+\sum_{l=0}^{\infty}\frac{2G}{3c^{2l+5}}\frac{1}{l!(2l+3)!!}\Big<\partial_{r}\hat{K}_{I_{l}}(t,r)\partial_{r}\hat{K}_{aI_{l}}(t,r)\Big>.
\end{align*}
Thus, substituting them in \eqref{equ4.38} and \eqref{equ4.39}, respectively, we obtain
\begin{align}
\label{equ4.51}\left(\frac{dE}{dt}\right)^{[0]}&=\left(\frac{dE}{dt}\right)^{[0]}_{GR}-\sum_{l=0}^{\infty}\frac{G}{3c^{2l+4}}\frac{1}{l!(2l+1)!!}\Big<\partial_{t}\hat{K}_{I_{l}}(t,r)\partial_{r}\hat{K}_{I_{l}}(t,r)\Big>,\\
\label{equ4.52}\left(\frac{dP_{a}}{dt}\right)^{[0]}&=\left(\frac{dP_{a}}{dt}\right)^{[0]}_{GR}+\sum_{l=0}^{\infty}\frac{2G}{3c^{2l+5}}\frac{1}{l!(2l+3)!!}\Big<\partial_{r}\hat{K}_{I_{l}}(t,r)\partial_{r}\hat{K}_{aI_{l}}(t,r)\Big>.
\end{align}
Finally, let us turn to dealing with the second term of (\ref{equ4.40}),
\begin{align}
\label{equ4.53}&+\frac{6r^3a^2}{\kappa}
\int\Big<\epsilon_{abc}n_{b}\partial_{r}R^{(1)}\partial_{c}R^{(1)}\Big>^{[3]}d\Omega\notag\\
=&+\frac{6r^3a^2}{\kappa}
\left(\int\Big<\epsilon_{abc}n_{b}\Big(\partial_{r}R^{(1)}\Big)^{[1]}\Big(\partial_{c}R^{(1)}\Big)^{[2]}\Big>d\Omega+
\int\Big<\epsilon_{abc}n_{b}\Big(\partial_{r}R^{(1)}\Big)^{[2]}\Big(\partial_{c}R^{(1)}\Big)^{[1]}\Big>d\Omega\right)\notag\\
=&+\frac{6r^3a^2}{\kappa}
\int\Big<\epsilon_{abc}n_{b}\Big(\partial_{r}\Big(R^{(1)}\Big)^{[1]}\Big)^{[1]}\Big(\partial_{c}R^{(1)}\Big)^{[2]}\Big>d\Omega\notag\\
=&+\frac{6r^3a^2}{\kappa}
\int\Big<\epsilon_{abc}n_{b}\Big(\partial_{r}\Big(R^{(1)}\Big)^{[1]}\Big)^{[1]}\Big(\Big(\partial_{c}\Big(R^{(1)}\Big)^{[1]}\Big)^{[2]}+\Big(\partial_{c}\Big(R^{(1)}\Big)^{[2]}\Big)^{[2]}\Big)\Big>d\Omega,
\end{align}
where in the second step, \eqref{equ4.32} and $\epsilon_{abc}n_{b}n_{c}=0$ have been used. Since $R^{(1)}=R^{(1)}(t,r,
\theta,\varphi)$,
\begin{align}
\label{equ4.54}\partial_{c}\Big(R^{(1)}\Big)^{[2]}&=(\partial_{c}r)\partial_{r}\Big(R^{(1)}\Big)^{[2]}
+(\partial_{c}\theta)\partial_{\theta}\Big(R^{(1)}\Big)^{[2]}+(\partial_{c}\varphi)\partial_{\varphi}\Big(R^{(1)}\Big)^{[2]}\notag\\
&=n_{c}\partial_{r}\Big(R^{(1)}\Big)^{[2]}
+(\partial_{c}\theta)\partial_{\theta}\Big(R^{(1)}\Big)^{[2]}+(\partial_{c}\varphi)\partial_{\varphi}\Big(R^{(1)}\Big)^{[2]}.
\end{align}
From \eqref{equ2.1}, there are
\begin{equation}\label{equ4.55}
\left\{\begin{array}{l}
\displaystyle \cos{\theta}=\frac{x_{3}}{\sqrt{x_{1}^{2}+x_{2}^{2}+x_{3}^{2}}},\medskip\\
\displaystyle \cos{\varphi}=\frac{x_{1}}{\sqrt{x_{1}^{2}+x_{2}^{2}}},\medskip\\
\displaystyle \sin{\varphi}=\frac{x_{2}}{\sqrt{x_{1}^{2}+x_{2}^{2}}},
\end{array}\right.
\end{equation}
which show easily the orders of $\partial_{c}\theta$, $\partial_{1}\varphi$, and $\partial_{2}\varphi$ are $1/r$, and further, $\partial_{3}\varphi=0$, so \eqref{equ4.54} implies
\begin{align}
\label{equ4.56}\Big(\partial_{c}\Big(R^{(1)}\Big)^{[2]}\Big)^{[2]}=n_{c}\Big(\partial_{r}\Big(R^{(1)}\Big)^{[2]}\Big)^{[2]}.
\end{align}
Therefore, \eqref{equ4.53} reduces to
\begin{align}
\label{equ4.57}&+\frac{6r^3a^2}{\kappa}
\int\Big<\epsilon_{abc}n_{b}\partial_{r}R^{(1)}\partial_{c}R^{(1)}\Big>^{[3]}d\Omega
=+\frac{6r^3a^2}{\kappa}
\int\Big<\epsilon_{abc}n_{b}\Big(\partial_{r}\Big(R^{(1)}\Big)^{[1]}\Big)^{[1]}\Big(\partial_{c}\Big(R^{(1)}\Big)^{[1]}\Big)^{[2]}\Big>d\Omega,
\end{align}
where $\epsilon_{abc}n_{b}n_{c}=0$ has been used.
From \eqref{equ3.35},
\begin{align}
\label{equ4.58}\partial_{c}\Big(R^{(1)}\Big)^{[1]}=&\frac{m^{2}\kappa}{4\pi r^2}n_{c}\sum_{l=0}^{\infty}\frac{1}{c^{l}l!}\hat{K}_{I_{l}}(t,r)N_{I_{l}}(\theta,\varphi)-\frac{m^{2}\kappa}{4\pi r}n_{c}\sum_{l=0}^{\infty}\frac{1}{c^{l}l!}\partial_{r}\hat{K}_{I_{l}}(t,r)N_{I_{l}}(\theta,\varphi)\notag\\
&-\frac{m^{2}\kappa}{4\pi r}\sum_{l=0}^{\infty}\sum_{p=1}^{l}\frac{1}{c^{l}l!}\hat{K}_{i_{p}I_{l-1}}(t,r)\big(\partial_{c}n_{i_{p}}\big)N_{I_{l-1}}(\theta,\varphi).
\end{align}
Then, by \eqref{equ2.13},
\begin{align}
\label{equ4.59}\Big(\partial_{c}\Big(R^{(1)}\Big)^{[1]}\Big)^{[2]}=&\frac{m^{2}\kappa}{4\pi r^2}n_{c}\sum_{l=0}^{\infty}\frac{1}{c^{l}l!}\hat{K}_{I_{l}}(t,r)N_{I_{l}}(\theta,\varphi)
-\frac{m^{2}\kappa}{4\pi r}\sum_{l=0}^{\infty}\sum_{p=1}^{l}\frac{1}{c^{l}l!}\hat{K}_{i_{p}I_{l-1}}(t,r)\big(\partial_{c}n_{i_{p}}\big)N_{I_{l-1}}(\theta,\varphi)\notag\\
=&-\frac{n_{c}}{r}\Big(R^{(1)}\Big)^{[1]}
-\frac{m^{2}\kappa}{4\pi r}\sum_{l=0}^{\infty}\sum_{p=1}^{l}\frac{1}{c^{l}l!}\hat{K}_{i_{p}I_{l-1}}(t,r)\big(\partial_{c}n_{i_{p}}\big)N_{I_{l-1}}(\theta,\varphi)\notag\\
=&\Big(\partial_{c}\Big(R^{(1)}\Big)^{[1]}\Big)^{[2]}_{\rm{eff}}+n_{c}\times\mbox{other terms}
\end{align}
with
\begin{align}
\label{equ4.60}
\Big(\partial_{c}\Big(R^{(1)}\Big)^{[1]}\Big)^{[2]}_{\rm{eff}}&=
-\frac{m^{2}\kappa}{4\pi r^2}\sum_{l=1}^{\infty}\frac{l}{c^{l}l!}\hat{K}_{cI_{l-1}}(t,r)N_{I_{l-1}}(\theta,\varphi).
\end{align}
Hence, by using $\epsilon_{abc}n_{b}n_{c}=0$ again,
\begin{align}
\label{equ4.61}&+\frac{6r^3a^2}{\kappa}
\int\Big<\epsilon_{abc}n_{b}\partial_{r}R^{(1)}\partial_{c}R^{(1)}\Big>^{[3]}d\Omega
=+\frac{6r^3a^2}{\kappa}
\int\Big<\epsilon_{abc}n_{b}\Big(\partial_{r}\Big(R^{(1)}\Big)^{[1]}\Big)^{[1]}\Big(\partial_{c}\Big(R^{(1)}\Big)^{[1]}\Big)^{[2]}_{\rm{eff}}\Big>d\Omega\notag\\
&=+\frac{6a^2}{\kappa}\bigg(\frac{m^{2}\kappa}{4\pi}\bigg)^2\sum_{l=0}^{\infty}
\sum_{l'=1}^{\infty}\frac{l'}{c^{l+l'}l!l'!}\int\Big<\epsilon_{abc}n_{b}\partial_{r}\hat{K}_{I_{l}}(t,r)N_{I_{l}}(\theta,\varphi)
\hat{K}_{cJ_{l'-1}}(t,r)N_{J_{l'-1}}(\theta,\varphi)\Big>d\Omega\notag\\
&=+\frac{G}{12\pi c^4}\sum_{l=0}^{\infty}\sum_{l'=1}^{\infty}\frac{l'}{c^{l+l'}l!l'!}\Big<\epsilon_{abc}\int\partial_{r}\hat{K}_{I_{l}}(t,r)N_{I_{l}}(\theta,\varphi)
\hat{K}_{cJ_{l'-1}}(t,r)N_{J_{l'-1}}(\theta,\varphi)n_{b}d\Omega\Big>,
\end{align}
where $m^2=1/6a$ and $\kappa=8\pi G/c^4$ have been used. Evaluating the integral in \eqref{equ4.61} with the help of the integral formula \eqref{equ2.19} gives
\begin{align*}
+\frac{6r^3a^2}{\kappa}
\int\Big<\epsilon_{abc}n_{b}\partial_{r}R^{(1)}\partial_{c}R^{(1)}\Big>^{[3]}d\Omega
&=+\frac{G}{12\pi c^4}\sum_{l=0}^{\infty}\sum_{l'=1}^{\infty}\frac{l'}{c^{l+l'}l!l'!}\left(
\Big<\epsilon_{abc}\partial_{r}\hat{K}_{bJ_{l'-1}}(t,r)
\hat{K}_{cJ_{l'-1}}(t,r)\Big>\frac{4\pi l'!}{(2l'+1)!!}\delta_{l\,l'}\right)\notag\\
&=+\sum_{l'=1}^{\infty}\frac{G}{3c^{2l'+4}}\frac{l'}{l'!(2l'+1)!!}\Big<\epsilon_{abc}\partial_{r}\hat{K}_{bJ_{l'-1}}(t,r)\hat{K}_{cJ_{l'-1}}(t,r)\Big>\notag\\
&=-\sum_{l=1}^{\infty}\frac{G}{3c^{2l+4}}\frac{1}{(l-1)!(2l+1)!!}\Big<\epsilon_{abc}\hat{K}_{bI_{l-1}}(t,r)\partial_{r}\hat{K}_{cI_{l-1}}(t,r)\Big>.
\end{align*}
Finally, substituting it in \eqref{equ4.40}, we obtain
\begin{align}
\label{equ4.62}\left(\frac{dJ_{a}}{dt}\right)^{[0]}&=\left(\frac{dJ_{a}}{dt}\right)^{[0]}_{GR}-\sum_{l=1}^{\infty}\frac{G}{3c^{2l+4}}\frac{1}{(l-1)!(2l+1)!!}\Big<\epsilon_{abc}\hat{K}_{bI_{l-1}}(t,r)\partial_{r}\hat{K}_{cI_{l-1}}(t,r)\Big>.
\end{align}

The above results show that due to the existence of the massive scalar degree of freedom, $R^{(1)}$, a gravitational system may have gravitational radiation in all multipoles, especially including the monopole and the dipole radiations.  The energy and the momentum in GWs are mainly contributed from their leading terms.  For the tensor part, the leading term is the quadrupole radiation, and for the scalar part, the leading term is the monopole radiation.  Eqs. (\ref{equ4.44}), (\ref{equ4.45}), (\ref{equ4.51}), and (\ref{equ4.52}) show that for the total power and rate of momentum, their quadrupole radiation in tensor part are proportional to $G/c^5$ and $G/c^7$, respectively, while the corresponding monopole radiation in scalar part are proportional to $G/(c^{4}v)$ and $G/(c^{5}v^2)$, for  $\partial_{r}$ in (\ref{equ4.51}) and (\ref{equ4.52}) can be rewritten as $\partial_{vt}$, where $v$ is the propagating velocity of scalar gravitational radiation.  In other words, there are $c/v$ and $c^2/v^2$ enhancements in the total power and rate of momentum for the monopole radiation compared with GR-like parts.  In contrast, Eq. (\ref{equ4.62}) shows that the monopole radiation does not contribute to the angular momentum, as expected.  The main correction to the total rate of angular momentum comes from the next leading term, namely, from the dipole radiation.  It will be much smaller than the angular momentum carried by a qradrupole radiation in the tensor part.  Therefore, for a given decay rate of energy of the source, the decay of momentum of the source in $f(R)$ gravity is more quickly than the GR counterpart, while the decay of angular momentum of the source in $f(R)$ gravity is slower than the GR counterpart.  Hence, the wave form of gravitational radiation in $f(R)$ gravity would be different from that in GR.  It also provides the criterion to test GR and sets the constraint on the coupling constant $a$ in (\ref{equ2.23}).
\section{Conclusions and discussions \label{Sec:Conclusion}}
It has been shown in this paper that the multipole expansion of linearized $f(R)$ gravity in the radiation field with irreducible Cartesian tensors is derived. Because the multipole expansion of linearized $f(R)$ gravity has been expressed in terms of the STF formalism \cite{Wu:2017huang},
by using the STF technique, it is feasible that the $1/r$-expansion in the distance to the source is applied to the linearized $f(R)$ gravity. The multipole moments associated with the tensor part of the source are only the functions of the retarded time $u=t-r/c$, as in GR, and however, the multipole moments associated with the scalar part~\cite{Wu:2017huang,Naf:2011za} of the source are the functions of $t$ and $r$ instead of the retarded time $u$.
This is foreseeable, because the tensor part $\tilde{h}^{\mu\nu}$ satisfies D'Alembert's equation, and represents the
propagation of the massless particle, but the scalar part $R^{(1)}$ satisfies Klein-Gordon equation, and represents the propagation of the massive particle.

Having obtained the multipole expansion of linearized $f(R)$ gravity in the radiation field, the energy, momentum, and angular momentum in the GWs can be dealt with.  With the help of the effective stress-energy of GWs in linearized $f(R)$ gravity, the energy and the momentum can be derived directly, but the angular momentum cannot, as in GR~\cite{Thorne:1980ru}. According to the Refs.~\cite{Thorne:1980ru,Peters:1964zz}, we find a method to make the energy, momentum, and angular momentum in the GWs for linearized $f(R)$ gravity dealt with in a unified way. This method requires the stress-energy pseudotensor of $f(R)$ gravity in the radiation field, instead of the effective stress-energy of GWs. Thus, we present the general expressions of the total power and rates of momentum and angular momentum in the GWs for linearized $f(R)$ gravity. These expressions have two parts which are associated with the tensor part $\tilde{h}^{\mu\nu}$ and the scalar part $R^{(1)}$ in the multipole expansion of linearized $f(R)$ gravity, respectively. The former are very the corresponding results in GR with the replace of the variables $h_{pq}^{TT}$ by $\tilde{h}_{pq}^{TT}$, and thus, the latter, from the massive scalar degree of freedom, are the corrections to the
results in GR.

As far as we know, such expressions in $f(R)$ gravity, especially for the momentum and the angular momentum, have not been given before. In Ref.~\cite{Naf:2011za}, the gravitational radiation in a quadratic $f(R)$ gravity has been investigated, and the total power in the GWs for the quadratic $f(R)$ has been obtained by a different way. But the total rates of momentum and angular momentum have not been calculated. With the help of the present results about the energy, momentum, and angular momentum in the GWs for $f(R)$ gravity and the data of observations, the coupling constant $a$ may be further constrained.  As mentioned in~\ref{I}, the GW generation formalism of $f(R)$ gravity based on the STF technique is worthy of being further investigated, and the multipole analysis about the radiation field in this paper is its important part.

\acknowledgments{This work is supported by the National Natural Science Foundation of China (Grant No.~11690022) and by the Strategic Priority Research Program of the Chinese Academy of Sciences "Multi-waveband Gravitational Wave
Universe" (Grant No. XDB23040000).}
\appendix\label{appendix}
\section{DERIVATION OF (\ref{equ4.29})}
In the derivation of \eqref{equ4.29}, besides the formulas \eqref{equ4.21}, \eqref{equ4.23}, and \eqref{equ4.26},
the following formulas are very useful~\cite{Peters:1964zz}:
\begin{align}
\label{equA1}\Big(\partial_{j}\partial_{\mu}\tilde{h}_{pq}^{TT}\Big)^{[2]}&=\Big(\partial_{j}\Big(\partial_{\mu}\tilde{h}_{pq}^{TT}\Big)^{[1]}\Big)^{[2]}
+\Big(\partial_{j}\Big(\partial_{\mu}\tilde{h}_{pq}^{TT}\Big)^{[2]}\Big)^{[2]},\\
\label{equA2}\Big(\partial_{j}\Big(\partial_{\mu}\tilde{h}_{pq}^{TT}\Big)^{[2]}\Big)^{[2]}&=-n_{j}\partial_{0}\Big(\partial_{\mu}\tilde{h}_{pq}^{TT}\Big)^{[2]}
=-n_{j}\Big(\partial_{0}\partial_{\mu}\tilde{h}_{pq}^{TT}\Big)^{[2]},\\
\label{equA3}\Big(\partial_{j}\Big(\tilde{h}_{ab}^{TT}\partial_{k}\tilde{h}_{pq}^{TT}\Big)^{[3]}\Big)^{[3]}&
=-n_{j}\partial_{0}\Big(\tilde{h}_{ab}^{TT}\partial_{k}\tilde{h}_{pq}^{TT}\Big)^{[3]}.
\end{align}
The $1/r^3$-terms in the average $\epsilon_{abc}n_{i}n_{b}(\tau^{ic}_{f})^{(2)}_{\rm{rad}}$ are,
\begin{align}
\label{equA4}\Big<\epsilon_{abc}n_{i}n_{b}(\tau^{ic}_{f})^{(2)}_{\rm{rad}}\Big>^{[3]}=&\frac{1}{2\kappa}\bigg(
-\Big<\epsilon_{abc}n_{i}n_{b}\tilde{h}_{pq}^{TT}\partial_{p}\partial_{q}\tilde{h}_{ic}^{TT}\Big>^{[3]}
+\frac{1}{2}\Big<\epsilon_{abc}n_{i}n_{b}\partial_{i}\tilde{h}_{pq}^{TT}\partial_{c}\tilde{h}_{pq}^{TT}\Big>^{[3]}
-\Big<\epsilon_{abc}n_{i}n_{b}\partial_{0}\tilde{h}_{ip}^{TT}\partial_{0}\tilde{h}_{cp}^{TT}\Big>^{[3]}\notag\\
&+\Big<\epsilon_{abc}n_{i}n_{b}\partial_{q}\tilde{h}_{ip}^{TT}\partial_{q}\tilde{h}_{cp}^{TT}\Big>^{[3]}
+\Big<\epsilon_{abc}n_{i}n_{b}\partial_{p}\tilde{h}_{iq}^{TT}\partial_{q}\tilde{h}_{cp}^{TT}\Big>^{[3]}
-\Big<\epsilon_{abc}n_{i}n_{b}\partial_{i}\tilde{h}_{pq}^{TT}\partial_{p}\tilde{h}_{cq}^{TT}\Big>^{[3]}\notag\\
&-\Big<\epsilon_{abc}n_{i}n_{b}\partial_{c}\tilde{h}_{pq}^{TT}\partial_{p}\tilde{h}_{iq}^{TT}\Big>^{[3]}
+12a^{2}\Big<\epsilon_{abc}n_{i}n_{b}\partial_{i}R^{(1)}\partial_{c}R^{(1)}\Big>^{[3]}\bigg).
\end{align}
The first term in \eqref{equA4} is
\begin{align}
\label{equA5}\Big<\epsilon_{abc}n_{i}n_{b}\tilde{h}_{pq}^{TT}\partial_{p}\partial_{q}\tilde{h}_{ic}^{TT}\Big>^{[3]}=&
\Big<\epsilon_{abc}n_{i}n_{b}\Big(\tilde{h}_{pq}^{TT}\Big)^{[1]}\Big(\partial_{p}\partial_{q}\tilde{h}_{ic}^{TT}\Big)^{[2]}\Big>
+\Big<\epsilon_{abc}n_{i}n_{b}\Big(\tilde{h}_{pq}^{TT}\Big)^{[2]}\Big(\partial_{p}\partial_{q}\tilde{h}_{ic}^{TT}\Big)^{[1]}\Big>.
\end{align}
The latter term is
\begin{align}
\label{equA6}&\Big<\epsilon_{abc}n_{i}n_{b}\Big(\tilde{h}_{pq}^{TT}\Big)^{[2]}\Big(\partial_{p}\partial_{q}\tilde{h}_{ic}^{TT}\Big)^{[1]}\Big>
=\Big<\epsilon_{abc}n_{i}n_{b}n_pn_q\Big(\tilde{h}_{pq}^{TT}\Big)^{[2]}\Big(\partial_{0}^2\tilde{h}_{ic}^{TT}\Big)^{[1]}\Big> \notag\\
=&\Big<\epsilon_{abc}n_{b}n_pn_q\Big(\tilde{h}_{pq}^{TT}\Big)^{[2]}\partial_{0}\Big(n_{i}\partial_{0}\tilde{h}_{ic}^{TT}\Big)^{[1]}\Big>
=-\Big<\epsilon_{abc}n_{b}n_pn_q\Big(\tilde{h}_{pq}^{TT}\Big)^{[2]}\partial_{0}\Big(\partial_{i}\tilde{h}_{ic}^{TT}\Big)^{[1]}\Big>=0.
\end{align}
In this derivation, (\ref{equ4.21}), (\ref{equ4.23}), and (\ref{equ4.26}) are used. So, combined with
(\ref{equ4.21}) and (\ref{equ4.23}) again, \eqref{equA5} reduces to
\begin{align}
\label{equA7}\Big<\epsilon_{abc}n_{i}n_{b}\tilde{h}_{pq}^{TT}\partial_{p}\partial_{q}\tilde{h}_{ic}^{TT}\Big>^{[3]}=
&\Big<\epsilon_{abc}n_{i}n_{b}\Big(\tilde{h}_{pq}^{TT}\Big)^{[1]}\Big(\partial_{p}(\partial_{q}\tilde{h}_{ic}^{TT})^{[1]}\Big)^{[2]}\Big>
+\Big<\epsilon_{abc}n_{i}n_{b}\Big(\tilde{h}_{pq}^{TT}\Big)^{[1]}\Big(\partial_{p}(\partial_{q}\tilde{h}_{ic}^{TT})^{[2]}\Big)^{[2]}\Big>\notag\\
=&-\Big<\epsilon_{abc}n_{i}n_{b}\Big(\tilde{h}_{pq}^{TT}\Big)^{[1]}\Big(\partial_{p}\Big(n_{q}\partial_{0}\tilde{h}_{ic}^{TT}\Big)^{[1]}\Big)^{[2]}\Big>
-\Big<\epsilon_{abc}n_{i}n_{b}n_{p}\Big(\tilde{h}_{pq}^{TT}\Big)^{[1]}\partial_{0}\Big(\partial_{q}\tilde{h}_{ic}^{TT}\Big)^{[2]}\Big>\notag\\
=&-\Big<\epsilon_{abc}n_{i}n_{b}\Big(\tilde{h}_{pq}^{TT}\Big)^{[1]}(\partial_{p}n_{q})\partial_{0}\Big(\tilde{h}_{ic}^{TT}\Big)^{[1]}\Big>
-\Big<\epsilon_{abc}n_{i}n_{b}n_{q}\Big(\tilde{h}_{pq}^{TT}\Big)^{[1]}\Big(\partial_{p}\partial_{0}\Big(\tilde{h}_{ic}^{TT}\Big)^{[1]}\Big)^{[2]}\Big>\notag\\
=&\Big<\epsilon_{abc}n_{b}\Big(\tilde{h}_{pq}^{TT}\Big)^{[1]}(\partial_{p}n_{q})\Big(\partial_{i}\tilde{h}_{ic}^{TT}\Big)^{[1]}\Big>
-\Big<\epsilon_{abc}n_{i}n_{b}n_{q}\Big(\tilde{h}_{pq}^{TT}\Big)^{[1]}\partial_{0}\Big(\partial_{p}\Big(\tilde{h}_{ic}^{TT}\Big)^{[1]}\Big)^{[2]}\Big>
=0.
\end{align}
In the first and second steps, (\ref{equA1}) and (\ref{equA2}) have been used, respectively. In the third and fifth steps,  (\ref{equ3.26}) has been used. On the similar reasons, the second term in \eqref{equA4} reads
\begin{align}
\label{equA8}\Big<\epsilon_{abc}n_{i}n_{b}\partial_{i}\tilde{h}_{pq}^{TT}\partial_{c}\tilde{h}_{pq}^{TT}\Big>^{[3]}=&
\Big<\epsilon_{abc}n_{i}n_{b}\Big(\partial_{i}\tilde{h}_{pq}^{TT}\Big)^{[1]}\Big(\partial_{c}\tilde{h}_{pq}^{TT}\Big)^{[2]}\Big>
+\Big<\epsilon_{abc}n_{i}n_{b}\Big(\partial_{i}\tilde{h}_{pq}^{TT}\Big)^{[2]}\Big(\partial_{c}\tilde{h}_{pq}^{TT}\Big)^{[1]}\Big>\notag\\
=&-\Big<\epsilon_{abc}n_{b}\Big(\partial_{0}\tilde{h}_{pq}^{TT}\Big)^{[1]}\Big(\partial_{c}\tilde{h}_{pq}^{TT}\Big)^{[2]}\Big>
-\Big<\epsilon_{abc}n_{i}n_{b}n_{c}\Big(\partial_{i}\tilde{h}_{pq}^{TT}\Big)^{[2]}\Big(\partial_{0}\tilde{h}_{pq}^{TT}\Big)^{[1]}\Big>\notag\\
=&-\Big<\epsilon_{abc}n_{b}\Big(\partial_{0}\tilde{h}_{pq}^{TT}\Big)^{[1]}\Big(\partial_{c}\tilde{h}_{pq}^{TT}\Big)^{[2]}\Big>
+\Big<\epsilon_{abc}n_{b}n_{c}\Big(\partial_{0}\tilde{h}_{pq}^{TT}\Big)^{[2]}\Big(\partial_{0}\tilde{h}_{pq}^{TT}\Big)^{[1]}\Big>\notag\\
=&-\Big<\epsilon_{abc}n_{b}\Big(\partial_{0}\tilde{h}_{pq}^{TT}\Big)^{[1]}\Big(\partial_{c}\tilde{h}_{pq}^{TT}\Big)^{[2]}\Big>
-\Big<\epsilon_{abc}n_{b}\Big(\partial_{0}\tilde{h}_{pq}^{TT}\Big)^{[2]}\Big(\partial_{c}\tilde{h}_{pq}^{TT}\Big)^{[1]}\Big>\notag\\
=&-\Big<\epsilon_{abc}n_{b}\partial_{0}\tilde{h}_{pq}^{TT}\partial_{c}\tilde{h}_{pq}^{TT}\Big>^{[3]},
\end{align}
where $\epsilon_{abc}n_{b}n_{c}=0$ has been used.
The fourth term in \eqref{equA4} is
\begin{align}
\label{equA9}\Big<\epsilon_{abc}n_{i}n_{b}\partial_{q}\tilde{h}_{ip}^{TT}\partial_{q}\tilde{h}_{cp}^{TT}\Big>^{[3]}
=&\Big<\epsilon_{abc}n_{i}n_{b}\Big(-\frac{1}{2}\big(\partial_{q}^{2}\tilde{h}_{ip}^{TT}\big)\tilde{h}_{cp}^{TT}
-\frac{1}{2}\tilde{h}_{ip}^{TT}\big(\partial_{q}^{2}\tilde{h}_{cp}^{TT}\big)
+\frac{1}{2}\partial_{q}\Big(\big(\partial_{q}\tilde{h}_{ip}^{TT}\big)\tilde{h}_{cp}^{TT}
+\tilde{h}_{ip}^{TT}\big(\partial_{q}\tilde{h}_{cp}^{TT}\big)\Big)\Big)\Big>^{[3]}.
\end{align}
The former two terms are
\begin{align}
\label{equA10}
\Big<\epsilon_{abc}n_{i}n_{b}\Big(-\frac{1}{2}\big(\partial_{q}^{2}\tilde{h}_{ip}^{TT}\big)\tilde{h}_{cp}^{TT}
-\frac{1}{2}\tilde{h}_{ip}^{TT}\big(\partial_{q}^{2}\tilde{h}_{cp}^{TT}\big)
\Big)\Big>^{[3]}=&\Big<\epsilon_{abc}n_{i}n_{b}\Big(-\frac{1}{2}\big(\partial_{0}^{2}\tilde{h}_{ip}^{TT}\big)\tilde{h}_{cp}^{TT}
-\frac{1}{2}\tilde{h}_{ip}^{TT}\big(\partial_{0}^{2}\tilde{h}_{cp}^{TT}\big)\Big)\Big>^{[3]}
\notag\\
=&\Big<\epsilon_{abc}n_{i}n_{b}\partial_{0}\tilde{h}_{ip}^{TT}\partial_{0}\tilde{h}_{cp}^{TT}\Big>^{[3]}.
\end{align}
In this derivation, the linearized gravitational field equations in the radiation field,
\begin{align}
\label{equA11}\partial_{0}^{2}\tilde{h}_{ij}^{TT}=\partial_{p}^{2}\tilde{h}_{ij}^{TT},
\end{align}
and Eq. \eqref{equ3.26} are used.  The latter two terms are
\begin{align}
\label{equA12}
&\Big<\epsilon_{abc}n_{i}n_{b}\Big(\frac{1}{2}\partial_{q}\Big(\big(\partial_{q}\tilde{h}_{ip}^{TT}\big)\tilde{h}_{cp}^{TT}
+\tilde{h}_{ip}^{TT}\big(\partial_{q}\tilde{h}_{cp}^{TT}\big)\Big)\Big)\Big>^{[3]}\notag\\
=&\frac{1}{2}\Big<\epsilon_{abc}n_{i}n_{b}\Big(\partial_{q}\Big(\big(\partial_{q}\tilde{h}_{ip}^{TT}\big)\tilde{h}_{cp}^{TT}
+\tilde{h}_{ip}^{TT}\big(\partial_{q}\tilde{h}_{cp}^{TT}\big)\Big)^{[2]}
+\partial_{q}\Big(\big(\partial_{q}\tilde{h}_{ip}^{TT}\big)\tilde{h}_{cp}^{TT}
+\tilde{h}_{ip}^{TT}\big(\partial_{q}\tilde{h}_{cp}^{TT}\big)\Big)^{[3]}\Big)\Big>^{[3]}\notag\\
=&\frac{1}{2}\Big<\epsilon_{abc}n_{i}n_{b}\Big(\partial_{q}\Big(\big(\partial_{q}\tilde{h}_{ip}^{TT}\big)\tilde{h}_{cp}^{TT}
+\tilde{h}_{ip}^{TT}\big(\partial_{q}\tilde{h}_{cp}^{TT}\big)\Big)^{[2]}
-n_{q}\partial_{0}\Big(\big(\partial_{q}\tilde{h}_{ip}^{TT}\big)\tilde{h}_{cp}^{TT}+\tilde{h}_{ip}^{TT}\big(\partial_{q}\tilde{h}_{cp}^{TT}\big)\Big)^{[3]}\Big)\Big>^{[3]}\notag\\
=&\frac{1}{2}\Big<\epsilon_{abc}n_{i}n_{b}\partial_{q}\Big(\big(\partial_{q}\tilde{h}_{ip}^{TT}\big)^{[1]}\big(\tilde{h}_{cp}^{TT}\big)^{[1]}
+\big(\tilde{h}_{ip}^{TT}\big)^{[1]}\big(\partial_{q}\tilde{h}_{cp}^{TT}\big)^{[1]}\Big)\Big>^{[3]}\notag\\
=&-\frac{1}{2}\Big<\epsilon_{abc}n_{i}n_{b}\partial_{q}\Big(n_{q}\big(\partial_{0}\tilde{h}_{ip}^{TT}\big)^{[1]}\big(\tilde{h}_{cp}^{TT}\big)^{[1]}
+\big(\tilde{h}_{ip}^{TT}\big)^{[1]}n_{q}\big(\partial_{0}\tilde{h}_{cp}^{TT}\big)^{[1]}\Big)\Big>^{[3]}\notag\\
=&-\frac{1}{2}\Big<\epsilon_{abc}n_{i}n_{b}\partial_{q}\Big(n_{q}\partial_{0}\big(\tilde{h}_{ip}^{TT}\tilde{h}_{cp}^{TT}\big)^{[2]}\Big)\Big>^{[3]}
=-\frac{1}{2}\Big<\partial_{0}\Big(\epsilon_{abc}n_{i}n_{b}
\partial_{q}\big(n_{q}\tilde{h}_{ip}^{TT}\tilde{h}_{cp}^{TT}\big)^{[2]}\Big)\Big>^{[3]}=0,
\end{align}
in which Eqs. \eqref{equ3.26} and \eqref{equA3} have been used.  Hence, the fourth term reads
\begin{align}
\label{equA13}\Big<\epsilon_{abc}n_{i}n_{b}\partial_{q}\tilde{h}_{ip}^{TT}\partial_{q}\tilde{h}_{cp}^{TT}\Big>^{[3]}
=\Big<\epsilon_{abc}n_{i}n_{b}\partial_{0}\tilde{h}_{ip}^{TT}\partial_{0}\tilde{h}_{cp}^{TT}\Big>^{[3]}.
\end{align}
The fifth term in \eqref{equA4} is
\begin{align}
\label{equA14}\Big<\epsilon_{abc}n_{i}n_{b}\partial_{p}\tilde{h}_{iq}^{TT}\partial_{q}\tilde{h}_{cp}^{TT}\Big>^{[3]}=&
\Big<\epsilon_{abc}n_{i}n_{b}\Big(\partial_{p}\tilde{h}_{iq}^{TT}\Big)^{[1]}\Big(\partial_{q}\tilde{h}_{cp}^{TT}\Big)^{[2]}\Big>
+\Big<\epsilon_{abc}n_{i}n_{b}\Big(\partial_{p}\tilde{h}_{iq}^{TT}\Big)^{[2]}\Big(\partial_{q}\tilde{h}_{cp}^{TT}\Big)^{[1]}\Big>.
\end{align}
The former term is
\begin{align}
\label{equA15} &\Big<\epsilon_{abc}n_{i}n_{b}\Big(\partial_{p}\tilde{h}_{iq}^{TT}\Big)^{[1]}\Big(\partial_{q}\tilde{h}_{cp}^{TT}\Big)^{[2]}\Big>
=-\Big<\epsilon_{abc}n_{i}n_{b}n_{p}\partial_{0}\Big(\tilde{h}_{iq}^{TT}\Big)^{[1]}\Big(\partial_{q}\tilde{h}_{cp}^{TT}\Big)^{[2]}\Big>\notag \\
=&\Big<\epsilon_{abc}n_{i}n_{b}n_{p}\Big(\tilde{h}_{iq}^{TT}\Big)^{[1]}\partial_{0}\Big(\partial_{q}\tilde{h}_{cp}^{TT}\Big)^{[2]}\Big>
=-\Big<\epsilon_{abc}n_{i}n_{b}\Big(\tilde{h}_{iq}^{TT}\Big)^{[1]}\Big(\partial_{p}\Big(\partial_{q}\tilde{h}_{cp}^{TT}\Big)^{[2]}\Big)^{[2]}\Big>\notag\\
=&-\Big<\epsilon_{abc}n_{i}n_{b}\Big(\tilde{h}_{iq}^{TT}\Big)^{[1]}\Big(\partial_{p}\partial_{q}\tilde{h}_{cp}^{TT}\Big)^{[2]}\Big>
+\Big<\epsilon_{abc}n_{i}n_{b}\Big(\tilde{h}_{iq}^{TT}\Big)^{[1]}\Big(\partial_{p}\Big(\partial_{q}\tilde{h}_{cp}^{TT}\Big)^{[1]}\Big)^{[2]}\Big>
\notag\\
=&-\Big<\epsilon_{abc}n_{i}n_{b}\Big(\tilde{h}_{iq}^{TT}\Big)^{[1]}\Big(\partial_{p}\Big(n_{q}\partial_{0}\tilde{h}_{cp}^{TT}\Big)^{[1]}\Big)^{[2]}\Big>
=-\Big<\epsilon_{abc}n_{i}n_{b}\Big(\tilde{h}_{iq}^{TT}\Big)^{[1]}\Big(\partial_{p}\Big(n_{q}\partial_{0}\Big(\tilde{h}_{cp}^{TT}\Big)^{[1]}\Big)\Big)^{[2]}\Big>
\notag \\
=&-\Big<\epsilon_{abc}n_{i}n_{b}\Big(\tilde{h}_{iq}^{TT}\Big)^{[1]}\big(\partial_{p}n_{q}\big)\Big(\partial_{0}\tilde{h}_{cp}^{TT}\Big)^{[1]}\Big>
-\Big<\epsilon_{abc}n_{i}n_{b}n_{q}\Big(\tilde{h}_{iq}^{TT}\Big)^{[1]}\partial_{0}\Big(\partial_{p}\Big(\tilde{h}_{cp}^{TT}\Big)^{[1]}\Big)^{[2]}\Big>\notag\\
=&-\frac{1}{r}\Big<\epsilon_{abc}n_{i}n_{b}\tilde{h}_{ip}^{TT}\partial_{0}\tilde{h}_{cp}^{TT}\Big>^{[2]}.
\end{align}
In the fourth step, (\ref{equA1}) is used.
Similarly, the latter term is
\begin{align}
\label{equA16}&\Big<\epsilon_{abc}n_{i}n_{b}\Big(\partial_{p}\tilde{h}_{iq}^{TT}\Big)^{[2]}\Big(\partial_{q}\tilde{h}_{cp}^{TT}\Big)^{[1]}\Big>
=-\Big<\epsilon_{abc}n_{i}n_{b}n_{q}\Big(\partial_{p}\tilde{h}_{iq}^{TT}\Big)^{[2]}\partial_{0}\Big(\tilde{h}_{cp}^{TT}\Big)^{[1]}\Big>\notag\\
=&\Big<\epsilon_{abc}n_{i}n_{b}n_{q}\partial_{0}\Big(\partial_{p}\tilde{h}_{iq}^{TT}\Big)^{[2]}\Big(\tilde{h}_{cp}^{TT}\Big)^{[1]}\Big>
=-\Big<\epsilon_{abc}n_{i}n_{b}\Big(\partial_{q}\Big(\partial_{p}\tilde{h}_{iq}^{TT}\Big)^{[2]}\Big)^{[2]}\Big(\tilde{h}_{cp}^{TT}\Big)^{[1]}\Big>\notag\\
=&\Big<\epsilon_{abc}n_{i}n_{b}\Big(\partial_{q}\Big(\partial_{p}\tilde{h}_{iq}^{TT}\Big)^{[1]}\Big)^{[2]}\Big(\tilde{h}_{cp}^{TT}\Big)^{[1]}\Big>
=-\Big<\epsilon_{abc}n_{i}n_{b}\Big(\partial_{q}\Big(n_{p}\partial_{0}\Big(\tilde{h}_{iq}^{TT}\Big)^{[1]}\Big)\Big)^{[2]}\Big(\tilde{h}_{cp}^{TT}\Big)^{[1]}\Big>\notag\\
=&-\Big<\epsilon_{abc}n_{i}n_{b}\big(\partial_{q}n_{p}\big)\Big(\partial_{0}\tilde{h}_{iq}^{TT}\Big)^{[1]}\Big(\tilde{h}_{cp}^{TT}\Big)^{[1]}\Big>
-\Big<\epsilon_{abc}n_{i}n_{b}n_{p}\partial_{0}\Big(\partial_{q}\Big(\tilde{h}_{iq}^{TT}\Big)^{[1]}\Big)^{[2]}\Big(\tilde{h}_{cp}^{TT}\Big)^{[1]}\Big>\notag\\
=&-\frac{1}{r}\Big<\epsilon_{abc}n_{i}n_{b}\partial_{0}\tilde{h}_{ip}^{TT}\tilde{h}_{cp}^{TT}\Big>^{[2]}.
\end{align}
So, the fifth term becomes
\begin{align}
\label{equA17}\Big<\epsilon_{abc}n_{i}n_{b}\partial_{p}\tilde{h}_{iq}^{TT}\partial_{q}\tilde{h}_{cp}^{TT}\Big>^{[3]}=&
-\frac{1}{r}\Big<\epsilon_{abc}n_{i}n_{b}\tilde{h}_{ip}^{TT}\partial_{0}\tilde{h}_{cp}^{TT}\Big>^{[2]}
-\frac{1}{r}\Big<\epsilon_{abc}n_{i}n_{b}\partial_{0}\tilde{h}_{ip}^{TT}\tilde{h}_{cp}^{TT}\Big>^{[2]}\notag\\
=&-\frac{1}{r}\Big<\partial_{0}\Big(\epsilon_{abc}n_{i}n_{b}\tilde{h}_{ip}^{TT}\tilde{h}_{cp}^{TT}\Big)\Big>^{[2]}=0.
\end{align}
The sixth term is
\begin{align}
\label{equA18}\Big<\epsilon_{abc}n_{i}n_{b}\partial_{i}\tilde{h}_{pq}^{TT}\partial_{p}\tilde{h}_{cq}^{TT}\Big>^{[3]}=&\Big<\epsilon_{abc}n_{i}n_{b}\Big(\partial_{i}\tilde{h}_{pq}^{TT}\Big)^{[1]}\Big(\partial_{p}\tilde{h}_{cq}^{TT}\Big)^{[2]}\Big>
+\Big<\epsilon_{abc}n_{i}n_{b}\Big(\partial_{i}\tilde{h}_{pq}^{TT}\Big)^{[2]}\Big(\partial_{p}\tilde{h}_{cq}^{TT}\Big)^{[1]}\Big>\notag\\
=&-\Big<\epsilon_{abc}n_{b}\Big(\partial_{0}\tilde{h}_{pq}^{TT}\Big)^{[1]}\Big(\partial_{p}\tilde{h}_{cq}^{TT}\Big)^{[2]}\Big>
-\Big<\epsilon_{abc}n_{i}n_{b}n_{p}\Big(\partial_{i}\tilde{h}_{pq}^{TT}\Big)^{[2]}\partial_{0}\Big(\tilde{h}_{cq}^{TT}\Big)^{[1]}\Big>.
\end{align}
The latter term is
\begin{align}
\label{equA19}&-\Big<\epsilon_{abc}n_{i}n_{b}n_{p}\Big(\partial_{i}\tilde{h}_{pq}^{TT}\Big)^{[2]}\partial_{0}\Big(\tilde{h}_{cq}^{TT}\Big)^{[1]}\Big>
=\Big<\epsilon_{abc}n_{i}n_{b}n_{p}\partial_{0}\Big(\partial_{i}\tilde{h}_{pq}^{TT}\Big)^{[2]}\Big(\tilde{h}_{cq}^{TT}\Big)^{[1]}\Big>\notag\\
=&-\Big<\epsilon_{abc}n_{i}n_{b}\Big(\partial_{p}\Big(\partial_{i}\tilde{h}_{pq}^{TT}\Big)^{[2]}\Big)^{[2]}\Big(\tilde{h}_{cq}^{TT}\Big)^{[1]}\Big>
=\Big<\epsilon_{abc}n_{i}n_{b}\Big(\partial_{p}\Big(\partial_{i}\tilde{h}_{pq}^{TT}\Big)^{[1]}\Big)^{[2]}\Big(\tilde{h}_{cq}^{TT}\Big)^{[1]}\Big>\notag\\
=&-\Big<\epsilon_{abc}n_{i}n_{b}\Big(\partial_{p}\Big(n_{i}\partial_{0}\Big(\tilde{h}_{pq}^{TT}\Big)^{[1]}\Big)\Big)^{[2]}\Big(\tilde{h}_{cq}^{TT}\Big)^{[1]}\Big>\notag\\
=&-\Big<\epsilon_{abc}n_{i}n_{b}\big(\partial_{p}n_{i}\big)\Big(\partial_{0}\tilde{h}_{pq}^{TT}\Big)^{[1]}\Big(\tilde{h}_{cq}^{TT}\Big)^{[1]}\Big>
-\Big<\epsilon_{abc}n_{b}\Big(\partial_{p}\partial_{0}\Big(\tilde{h}_{pq}^{TT}\Big)^{[1]}\Big)^{[2]}\Big(\tilde{h}_{cq}^{TT}\Big)^{[1]}\Big>\notag\\
=&-\frac{1}{r}\Big<\epsilon_{abc}n_{i}n_{b}\Big(\partial_{0}\tilde{h}_{iq}^{TT}\Big)^{[1]}\Big(\tilde{h}_{cq}^{TT}\Big)^{[1]}\Big>
-\frac{1}{r}\Big<\epsilon_{abc}n_{b}\Big(\partial_{p}\tilde{h}_{pq}^{TT}\Big)^{[1]}\Big(\tilde{h}_{cq}^{TT}\Big)^{[1]}\Big>\notag\\
&-\Big<\epsilon_{abc}n_{b}\Big(\partial_{p}\Big(\partial_{0}\tilde{h}_{pq}^{TT}\Big)^{[1]}\Big)^{[2]}\Big(\tilde{h}_{cq}^{TT}\Big)^{[1]}\Big>\notag\\
=&-\frac{1}{r}\Big<\epsilon_{abc}n_{i}n_{b}\partial_{0}\tilde{h}_{iq}^{TT}\tilde{h}_{cq}^{TT}\Big>^{[2]}
+\Big<\epsilon_{abc}n_{b}\Big(\partial_{p}\Big(\partial_{0}\tilde{h}_{pq}^{TT}\Big)^{[2]}\Big)^{[2]}\Big(\tilde{h}_{cq}^{TT}\Big)^{[1]}\Big>\notag\\
=&-\frac{1}{r}\Big<\epsilon_{abc}n_{i}n_{b}\partial_{0}\tilde{h}_{iq}^{TT}\tilde{h}_{cq}^{TT}\Big>^{[2]}
-\Big<\epsilon_{abc}n_{b}n_{p}\partial_{0}\Big(\partial_{0}\tilde{h}_{pq}^{TT}\Big)^{[2]}\Big(\tilde{h}_{cq}^{TT}\Big)^{[1]}\Big>\notag\\
=&-\frac{1}{r}\Big<\epsilon_{abc}n_{i}n_{b}\partial_{0}\tilde{h}_{iq}^{TT}\tilde{h}_{cq}^{TT}\Big>^{[2]}
+\Big<\epsilon_{abc}n_{b}n_{p}\Big(\partial_{0}\tilde{h}_{pq}^{TT}\Big)^{[2]}\Big(\partial_{0}\tilde{h}_{cq}^{TT}\Big)^{[1]}\Big>\notag\\
=&-\frac{1}{r}\Big<\epsilon_{abc}n_{i}n_{b}\partial_{0}\tilde{h}_{iq}^{TT}\tilde{h}_{cq}^{TT}\Big>^{[2]}
-\Big<\epsilon_{abc}n_{b}\Big(\partial_{0}\tilde{h}_{pq}^{TT}\Big)^{[2]}\Big(\partial_{p}\tilde{h}_{cq}^{TT}\Big)^{[1]}\Big>.
\end{align}
Substituting \eqref{equA19} in \eqref{equA18}, we obtain
\begin{align}
\label{equA20}\Big<\epsilon_{abc}n_{i}n_{b}\partial_{i}\tilde{h}_{pq}^{TT}\partial_{p}\tilde{h}_{cq}^{TT}\Big>^{[3]}
=&-\Big<\epsilon_{abc}n_{b}\partial_{0}\tilde{h}_{pq}^{TT}\partial_{p}\tilde{h}_{cq}^{TT}\Big>^{[3]}
-\frac{1}{r}\Big<\epsilon_{abc}n_{i}n_{b}\partial_{0}\tilde{h}_{iq}^{TT}\tilde{h}_{cq}^{TT}\Big>^{[2]}.
\end{align}
The seventh term is
\begin{align}
\label{equA21}\Big<\epsilon_{abc}n_{i}n_{b}\partial_{c}\tilde{h}_{pq}^{TT}\partial_{p}\tilde{h}_{iq}^{TT}\Big>^{[3]}=&
\Big<\epsilon_{abc}n_{i}n_{b}\Big(\partial_{c}\tilde{h}_{pq}^{TT}\Big)^{[1]}\Big(\partial_{p}\tilde{h}_{iq}^{TT}\Big)^{[2]}\Big>
+\Big<\epsilon_{abc}n_{i}n_{b}\Big(\partial_{c}\tilde{h}_{pq}^{TT}\Big)^{[2]}\Big(\partial_{p}\tilde{h}_{iq}^{TT}\Big)^{[1]}\Big>\notag\\
=&-\Big<\epsilon_{abc}n_{i}n_{b}n_{c}\Big(\partial_{0}\tilde{h}_{pq}^{TT}\Big)^{[1]}\Big(\partial_{p}\tilde{h}_{iq}^{TT}\Big)^{[2]}\Big>
-\Big<\epsilon_{abc}n_{i}n_{b}n_{p}\Big(\partial_{c}\tilde{h}_{pq}^{TT}\Big)^{[2]}\Big(\partial_{0}\tilde{h}_{iq}^{TT}\Big)^{[1]}\Big>\notag\\
=&-\Big<\epsilon_{abc}n_{i}n_{b}n_{p}\Big(\partial_{c}\tilde{h}_{pq}^{TT}\Big)^{[2]}\partial_{0}\Big(\tilde{h}_{iq}^{TT}\Big)^{[1]}\Big>
=\Big<\epsilon_{abc}n_{i}n_{b}n_{p}\partial_{0}\Big(\partial_{c}\tilde{h}_{pq}^{TT}\Big)^{[2]}\Big(\tilde{h}_{iq}^{TT}\Big)^{[1]}\Big>\notag\\
=&-\Big<\epsilon_{abc}n_{i}n_{b}\Big(\partial_{p}\Big(\partial_{c}\tilde{h}_{pq}^{TT}\Big)^{[2]}\Big)^{[2]}\Big(\tilde{h}_{iq}^{TT}\Big)^{[1]}\Big>
=\Big<\epsilon_{abc}n_{i}n_{b}\Big(\partial_{p}\Big(\partial_{c}\tilde{h}_{pq}^{TT}\Big)^{[1]}\Big)^{[2]}\Big(\tilde{h}_{iq}^{TT}\Big)^{[1]}\Big>\notag\\
=&-\Big<\epsilon_{abc}n_{i}n_{b}\Big(\partial_{p}\Big(n_{c}\partial_{0}\tilde{h}_{pq}^{TT}\Big)^{[1]}\Big)^{[2]}\Big(\tilde{h}_{iq}^{TT}\Big)^{[1]}\Big>
\notag\\
=&-\Big<\epsilon_{abc}n_{i}n_{b}\big(\partial_{p}n_{c}\big)\Big(\partial_{0}\tilde{h}_{pq}^{TT}\Big)^{[1]}\Big(\tilde{h}_{iq}^{TT}\Big)^{[1]}\Big>
-\Big<\epsilon_{abc}n_{i}n_{b}n_{c}\Big(\partial_{p}\Big(\partial_{0}\tilde{h}_{pq}^{TT}\Big)^{[1]}\Big)^{[2]}\Big(\tilde{h}_{iq}^{TT}\Big)^{[1]}\Big>\notag\\
=&-\frac{1}{r}\Big<\epsilon_{abc}n_{i}n_{b}\Big(\partial_{0}\tilde{h}_{cq}^{TT}\Big)^{[1]}\Big(\tilde{h}_{iq}^{TT}\Big)^{[1]}\Big>
+\frac{1}{r}\Big<\epsilon_{abc}n_{i}n_{b}n_{c}n_{p}\Big(\partial_{0}\tilde{h}_{cq}^{TT}\Big)^{[1]}\Big(\tilde{h}_{iq}^{TT}\Big)^{[1]}\Big>\notag\\
=&-\frac{1}{r}\Big<\epsilon_{abc}n_{i}n_{b}\tilde{h}_{iq}^{TT}\partial_{0}\tilde{h}_{cq}^{TT}\Big>^{[2]}.
\end{align}
Therefore, Eq.~\eqref{equA4} reads
\begin{align}
\label{equA22}\Big<\epsilon_{abc}n_{i}n_{b}(\tau^{ic}_{f})^{(2)}_{\rm{rad}}\Big>^{[3]}=&\frac{1}{2\kappa}\Big(
-\frac{1}{2}\Big<\epsilon_{abc}n_{b}\partial_{0}\tilde{h}_{pq}^{TT}\partial_{c}\tilde{h}_{pq}^{TT}\Big>^{[3]}
-\Big<\epsilon_{abc}n_{i}n_{b}\partial_{0}\tilde{h}_{ip}^{TT}\partial_{0}\tilde{h}_{cp}^{TT}\Big>^{[3]}\notag\\
&+\Big<\epsilon_{abc}n_{i}n_{b}\partial_{0}\tilde{h}_{ip}^{TT}\partial_{0}\tilde{h}_{cp}^{TT}\Big>^{[3]}
+\Big<\epsilon_{abc}n_{b}\partial_{0}\tilde{h}_{pq}^{TT}\partial_{p}\tilde{h}_{cq}^{TT}\Big>^{[3]}
+\frac{1}{r}\Big<\epsilon_{abc}n_{i}n_{b}\partial_{0}\tilde{h}_{iq}^{TT}\tilde{h}_{cq}^{TT}\Big>^{[2]}\notag\\
&+\frac{1}{r}\Big<\epsilon_{abc}n_{i}n_{b}\tilde{h}_{iq}^{TT}\partial_{0}\tilde{h}_{cq}^{TT}\Big>^{[2]}
+12a^{2}\Big<\epsilon_{abc}n_{i}n_{b}\partial_{i}R^{(1)}\partial_{c}R^{(1)}\Big>^{[3]}\Big) \notag \\
=&\frac{1}{2\kappa}\Big(\Big<\epsilon_{abc}n_{b}\partial_{0}\tilde{h}_{pq}^{TT}\partial_{p}\tilde{h}_{cq}^{TT}\Big>^{[3]}
-\frac{1}{2}\Big<\epsilon_{abc}n_{b}\partial_{0}\tilde{h}_{pq}^{TT}\partial_{c}\tilde{h}_{pq}^{TT}\Big>^{[3]}
+12a^{2}\Big<\epsilon_{abc}n_{b}\partial_{r}R^{(1)}\partial_{c}R^{(1)}\Big>^{[3]}\Big) \notag\\
=&\frac{1}{2\kappa}\Big(\Big<\epsilon_{abc}n_{b}\Big(\partial_{p}\Big(\tilde{h}_{cq}^{TT}\partial_{0}\tilde{h}_{pq}^{TT}\Big)
-\frac{1}{2}\partial_{0}\tilde{h}_{pq}^{TT}\partial_{c}\tilde{h}_{pq}^{TT}\Big)\Big>^{[3]}
+12a^{2}\Big<\epsilon_{abc}n_{b}\partial_{r}R^{(1)}\partial_{c}R^{(1)}\Big>^{[3]}\Big).
\end{align}
By using (\ref{equ3.26}) again, we acquire \eqref{equ4.29}.

\end{document}